\documentclass[sigconf]{acmart}

\usepackage{amsmath}
\usepackage{multicol}
\usepackage{multirow}

\usepackage{siunitx}
\usepackage{tabularx}
\usepackage[shortlabels]{enumitem}
\usepackage{array}
\usepackage{makecell}

\AtBeginDocument{%
  \providecommand\BibTeX{{%
    \normalfont B\kern-0.5em{\scshape i\kern-0.25em b}\kern-0.8em\TeX}}}

\setcopyright{acmlicensed}
\copyrightyear{2025}
\acmYear{2025}
\setcopyright{acmlicensed}
\acmConference[CHI '25]{CHI Conference on Human Factors in Computing Systems}{April 26-May 1, 2025}{Yokohama, Japan} \acmBooktitle{CHI Conference on Human Factors in Computing Systems (CHI '25), April 26-May 1, 2025, Yokohama, Japan} \acmDOI{10.1145/3706598.3713882}
\acmISBN{979-8-4007-1394-1/2025/04}

\newcommand{\engquote}[1]{\textit{``#1''}}

\usepackage{multicol}
\usepackage{multirow}
\usepackage{siunitx}
\usepackage{subcaption}
\usepackage{tabularx}
\usepackage{svg}

\def\markup{0}

\if\markup1
\newcommand{\revision}[1]{{\leavevmode\color{blue}#1}}
\usepackage{soul}
\usepackage[normalem]{ulem}
\else
\newcommand{\revision}[1]{#1}
\newcommand{\st}[1]{}
\newcommand{\sout}[1]{}
\fi

\begin{document}



\title[Interactive Digital Item]{InteRecon: Towards Reconstructing Interactivity of Personal Memorable Items in Mixed Reality}

\author{Zisu Li}
\orcid{0000-0001-8825-0191}
\email{zlihe@connect.ust.hk}
\affiliation{%
  \institution{The Hong Kong University of Science and Technology}
  \city{Hong Kong SAR}
  \country{China}}
\affiliation{%
  \institution{MIT CSAIL}
  \country{Cambridge, MA, USA}
  }

\author{Jiawei Li}
\email{jli526@connect.hkust-gz.edu.cn}
\orcid{0009-0000-6593-2958}
\affiliation{
  \institution{The Hong Kong University of Science and Technology (Guangzhou)}
  \city{Guangzhou}
  \country{China}
}

\author{Zeyu Xiong}
\email{zeyu.xiong@inf.ethz.ch}
\orcid{0000-0002-3652-1890}
\affiliation{
  \institution{Department of Computer Science ETH Zurich}
  \country{Zürich, Switzerland}
}

\author{Shumeng Zhang}
\email{szhang390@connect.hkust-gz.edu.cn}
\orcid{0009-0005-2000-6113}
\affiliation{
  \institution{The Hong Kong University of Science and Technology (Guangzhou)}
  \country{Guangzhou, China}
}

\author{Faraz Faruqi}
\email{ffaruqi@mit.edu}
\orcid{0000-0002-1691-2093}
\affiliation{
  \institution{MIT CSAIL}
  \country{Cambridge, MA, USA}
}

\author{Stefanie Mueller}
\email{stefanie.mueller@mit.edu}
\orcid{0000-0001-7743-7807}
\affiliation{
  \institution{MIT CSAIL}
  \country{Cambridge, MA, USA}
}



\author{Chen Liang}
\email{chenliang2@hkust-gz.edu.cn}
\orcid{0000-0003-0579-2716}
\affiliation{
  \institution{The Hong Kong University of Science and Technology (Guangzhou)}
  \city{Guangzhou}
  \country{China}
}

\author{Xiaojuan Ma}
\email{mxj@cse.ust.hk}
\orcid{0000-0002-9847-7784}
\affiliation{
  \institution{The Hong Kong University of Science and Technology}
  \city{Hong Kong SAR}
  \country{China}
  }

\author{Mingming Fan}
\email{mingmingfan@ust.hk}
\orcid{0000-0002-0356-4712}
\affiliation{
  \institution{The Hong Kong University of Science and Technology (Guangzhou)}
  \city{Guangzhou}
  \country{China}
  }
\affiliation{
  \institution{The Hong Kong University of Science and Technology}
  \city{Hong Kong SAR}
  \country{China}
}


\renewcommand{\shortauthors}{Li et al.}

\begin{abstract}

Digital capturing of memorable personal items is a key way to archive personal memories. Although current digitization methods (e.g., photos, videos, 3D scanning) can replicate the physical appearance of an item, they often cannot preserve its real-world interactivity. We present \textit{Interactive Digital Item (IDI)}, a concept of reconstructing both the physical appearance and, more importantly, the interactivity of an item. We first conducted a formative study to understand users' expectations of IDI, identifying key physical interactivity features, including geometry, interfaces, and embedded content of items. Informed by these findings, we developed \textit{InteRecon}, an AR prototype enabling personal reconstruction functions for IDI creation. An exploratory study was conducted to assess the feasibility of using InteRecon and explore the potential of IDI to enrich personal memory archives. 
Results show that InteRecon is feasible for IDI creation, and the concept of IDI brings new opportunities for augmenting personal memory archives.



\end{abstract}


 \begin{teaserfigure}
   \includegraphics[width=\linewidth]{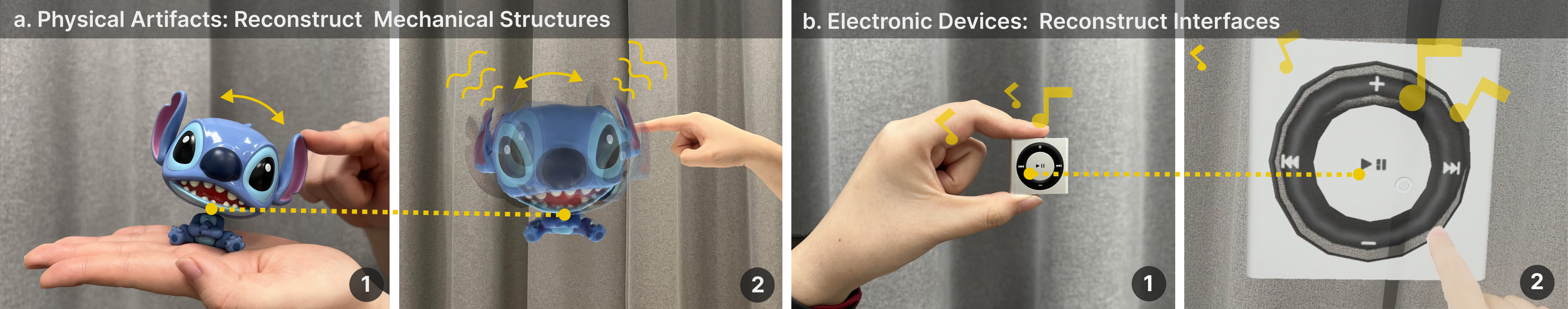}
   \caption{\textbf{Two examples of reconstructing memorable items while preserving their interactivity: (a-1) Touching the toy Stitch and making its head shake as in the real world, (a-2) Touching the reconstructed toy Stitch in AR to achieve similar motions as the real world by InteRecon to reconstruct its mechanical structures, (b-1) Using a physical iPod in the real world, (b-2) Using the reconstructed iPod to play music in AR by InteRecon to add corresponding functions to all the buttons in the interface.}}
   \Description{Figure 1 shows two examples of reconstructing memorable items with preserving their interactivity: (a-1) Touching the toy Stitch and making its head shake in the real world, (a-2) Touching the reconstructed toy Stitch in AR to achieve similar motions with the real world by using InteRecon to reconstruct its mechanical structures, (b-1) Using a physical iPod in the real world, (b-2) Using the reconstructed iPod to play music in AR by adding corresponding functions to all the buttons in the interface.}
   \label{fig:teaser}
 \end{teaserfigure}


\maketitle
    \section{Introduction}

Memorable personal items function as catalysts, invoking and symbolizing significant locations, moments, objects, individuals, and experiences to represent memory phases \cite{10.1145/1806923.1806924,petrelli2010family, bowen2011remembering}. Humans seek lifelong preservation of personal memory, mainly through the digitization of physical content.
Current digitization methods for personal memories are largely based on videos or pictures, with an auxiliary focus on capturing textual descriptions and physical appearances of memorable items \cite{petrelli2010family,KALNIKAITE2011298}. 
However, digital copies created by these methods are often unable to fully express the richness of memories, as they do not inherit the interactive features (e.g., the physical, sensory, or functional use) from the physical items. 
Imagine that using your grandmother’s jam book or soup ladle occasionally ignites a story your mother told about your grandmother---a digital copy of videos or pictures could never be used in daily life \cite{10.1145/1806923.1806924}.
Moreover, prior work found the stories of past memories associated with memorable items may be `stumbled across' in people's interactions or functional use of physical objects in the real world, while digital photos or videos tend to be stored away and do not become integrated into people's real life \cite{petrelli2010family,KALNIKAITE2011298,west2007memento}.
\revision{Therefore, personal reconstructing the interactivity of memorable items in their digital counterparts is valuable in facilitating personal memory archives in people's daily lives.}

In this paper, we present \textit{Interactive Digital Item} (IDI), a novel concept for personal item digital reconstruction featuring the preservation of its physical interactivity. 
The concept of IDI extends beyond the conventional `interactive 3D models' typically created by developers or modelers. IDI includes an authoring process that is as accessible as photo capturing, allowing non-experts to contribute to enriching personal memory archives.
We started with a formative study to investigate users' expectations of the essential attributes that constitute the physical interactivity of memorable items.
Two prominent types of memorable items are proposed by participants as most required for reconstruction - obsolete electronic devices (e.g., music players, cassette players, game controllers) and physical artifacts (e.g., dairies, albums, toys) - because users found the interactive features in these items largely convey the traces of memories.
Three design goals were devised from the study that needed to be incorporated into IDI, in which IDI should: 1) be a reconstructed 3D model from the physical item with similar visual properties (e.g., size, shape, texture) and physical properties (e.g., gravity, collisions, motions), 2) demonstrate the original interactivity within the interfaces of electronic devices by activating the tangible widgets (e.g., pressing buttons, dragging sliders) on devices and 3) preserve the embedded content of digital files (e.g., photos, songs, or software) in electronic devices and the usage scenario contexts for physical artifacts.


Based on IDI's design goals, we implemented an AR prototype, namely \textit{InteRecon}, to support the end-user creation of IDI from a physical item in the Mixed Reality (MR) environment. 
InteRecon provides four core functions - reconstructing 3D appearance, adding physical transforms, reconstructing interface, and adding embedded content - for physical interactivity reconstruction. 
We conducted a two-session user study with 16 participants to understand the feasibility of using InteRecon to create IDI and further explore the participants’ experiences of using InteRecon to reconstruct their own items, collecting feedback on the challenges and future opportunities of IDI.
Results show that InteRecon was effective and enjoyable for IDI creation, and the physical interactivity created by participants augmented the realistic experiences of participants. Moreover, participants also proposed that creative reconstructions go beyond real-world interactivity, along with in-situ and life-logging usage scenarios of IDIs, to enrich memory archives. 
We make the following contributions:
\begin{itemize}
    \item A novel concept of IDI derived from a formative study, a digital personalized reconstruction of memorable personal items while maintaining their physical interactivity features.
    \item An AR prototype, InteRecon, enables end-users to create IDI incorporating design goals of reconstructing geometry, interface, and embedded content.
    \item Potential opportunities and applications of IDI discovered from a the user study in terms of realism creation, interactivity approaches, and customization potentials. Future avenues are also discussed to augment the creation and usage of IDI for enriching personal memory archives. 
\end{itemize}

    \section{Related Work}
\subsection{Real-world Interactive Features of Memorable Items}

Previous research has explored the characteristics of memorable personal items that activate recollections. 
Physical objects, along with the environment or context in which a remembered activity occurred, play an important role in how an activity is represented in one's memory \cite{10.1145/2702613.2732842,kawamura2007ubiquitous,li2020facilitating,10.1145/3027063.3052756}. 
The tangible nature of physical objects, including elements you can visually perceive, touch, or interact with, are important in understanding the essence of memory activities \cite{10.1145/1394445.1394472,10.1145/2702613.2732842,habermas2002souvenirs,marschall2019memory,10.1145/2470654.2466453}. 
For instance, the physical interactions involved in an activity, such as the tools utilized, the space it occurs in, and other physical aspects, can create enduring memories \cite{10.1145/1806923.1806924}.
These memories can easily resurface in everyday life, particularly when one incidentally encounters them through the functional use or the random interactions of a physical object \cite{petrelli2010family,KALNIKAITE2011298,west2007memento}.  
\citet{10.1145/3024969.3024996} mentioned that tangible real-world interaction such as flipping pages or arranging blocks can improve concentration in collocated communication about memories and even stimulate collaboration in memory activities. 
\citet{10.1145/1394445.1394472} found people favored using physical souvenirs from travel to access photo sets, appreciating the serendipitous sharing of physical souvenirs.
\citet{10.1145/1517664.1517678} investigated the advantages of sharing family photographs facilitated the physical affordances of objects (e.g., the ability to store and display images) within the home environment, integrating them into daily routines.
Additionally, there is a growing focus on how physical interactions that can open up new ways to organize and collective memories, provoke social conversations \cite{hawkins2015postulater,hilliges2009getting,10.1145/3332165.3347877}, and support collective experiences of reminiscence and reflection \cite{hawkins2015postulater,jansen2014pearl,10.1145/2317956.2318055,10.1145/1753326.1753635,10.1145/2598510.2598589,10.1145/1394445.1394461,10.1145/3332165.3347877}.

Although existing work has incorporated different interactive features into personal items to improve usability and immersiveness, none of them has explicitly revealed the user expectations regarding the indispensable attributes of physical interactivity for activating personal memories.
In this paper, we bridge this gap by conducting a formative study to identify essential physical interactivity attributes of memorable personal items, with the goal of reconstructing memorable personal items with interactivity integrated into the digital realm to augment the preservation of personal memories. 

\subsection{Digital and Physical Forms of Personal Memory Archives}
Two primitive forms of digital media used for personal memory archives are photographs and videos. 
To facilitate efficient content creation~\cite{mayer2009establishing, somerville2011recuerdos, dobbins2014creating}, organization~\cite{neumayer2005content, runardotter2009organizing}, and access~\cite{addis2010100, beigl2001mediacups} processes for personal memory archives, previous research has investigated novel digital designs by extending the original photographs and video forms~\cite{garde2011digital, potter2010embodied, mezaris2018personal, hoskins2017restless,10.1145/3532106.3533501}.
For example, \textit{Rewind}~\cite{10.1145/3287069} proposed to associate one's daily excursion videos with a sequence of street-level images from self-tracked location data, aiming to organize video contents in a geographical for better visualization and access.
\textit{Chronoscope} is domestic technology leveraging temporal metadata in digital photos as a resource to encourage diverse and open-ended experiences when revisiting one's personal digital photo archive \cite{10.1145/3322276.3322301}. 
\emph{Shoebox} offers a digital solution for combining the storage and display of digital images within the home environment \cite{10.1145/1517664.1517678}.

Although the above digital capturing methods may offer visual accuracy and immediate representational value, they often lack the symbolic and emotional meanings inherent to physical objects associated with memory experiences \cite{petrelli2010family,kirk2010human,jones2018co}.
These meanings are essential for enhancing personal memory archives and have a lasting impact beyond the immediate representational value. \cite{bradley2014emotional,jones2018co,cushing2011self}. 
Research has extensively compared the impact of physical objects versus digital photos and videos on memory activation, consistently finding a preference for physical objects \cite{10.1145/2317956.2318054,petrelli2010family,10.1145/2470654.2466453}. 
Physical objects, with their tangible presence, allow for direct physical interaction, capturing the essence of experiences more profoundly than digital alternatives \cite{kirk2010human,nunes2009using,petrelli2014photo}.
For instance, Petrelli et al. \cite{petrelli2010family,petrelli2008autotopography} found physical mementos are highly valued heterogeneous and support different types of recollection compared to digital photos and videos.
\citet{10.1145/1031607.1031673} discovered that it is more common for participants to actively share memories with physical items than digital artifacts such as photos and videos when they are outside the home. 
While physical items play a crucial role in triggering memories, they have disadvantages, such as fragility, degradation, and the need for additional storage space, which increases costs \cite{kirk2010human}. 
To address these challenges, digital preservation becomes important to ensure the long-term safeguarding of these items. Despite their importance, the interactive attributes of physical items have rarely been considered in the digital reconstruction of memory archives.

Our work aligns with previous efforts to achieve digital longevity, ensure the persistence of memories, and facilitate the efficient evocation of memories and emotions \cite{duranti2012memory}. 
However, unlike earlier approaches, we enrich the concept of personal digital reconstruction by allowing end-users to integrate physical interactivity into digitally reconstructed items. Additionally, we explore the design space of this reconstruction process across three dimensions: geometry, interface, and embedded content.

\subsection{Immersive Technologies to Facilitate Memory Reconstruction}
Memory reconstruction is a process where the user records memory elements, archives them in digital forms, and revisits them for recollection~\cite{pollio1971memory, crete2012reconstructing, maddali2023understanding}. 
Immersive technologies, typically built upon AR and VR platforms, play an important role in assisting memory reconstruction in different stages \cite{valtolina2005dissemination,10.1145/3610903,iriye2021memories}.
Their immersive display and interaction capabilities stand out as key advantages for memory reconstruction, while the mobility of these devices further enhances their utility in common memory-sharing and communication scenarios \cite{10.1145/3024969.3024996,kirk2010human}.

With high-fidelity 3D display technologies, AR or VR could create more immersive and realistic representations of memories \cite{lekan2016virtual,10.1145/3385956.3418945,krokos2019virtual}.
They thus provide historians an invaluable tool for the realistic reconstruction of ancient artifacts, past heritages, and memories, giving them more longevity within the digital realm \cite{tan2009virtual}.
For example, Valtolina et al. \cite{valtolina2005dissemination} and Yang et al. \cite{yang2006creating} were devoted to reconstructing ancient sites in the form of accurate 3D models in mixed reality, enhancing the visitors' understanding, accessibility, and engagement with historical narratives.
\citet{10.1145/3365610.3368425} leveraged VR to create a virtual graveyard as an accurate simulation of the Salla graveyard as possible even with its atmosphere, providing a deeply immersive experience for visitors.


The interaction methods offered by immersive technologies can provide more intuitive and flexible operations for 3D content (e.g., models, scenarios, etc) for non-expert users\cite{10.1145/3472749.3474769,whitlock2020mrcat,10.1145/3544548.3581148}. 
For example, \textit{GesturAR} is an end-to-end authoring tool that supports users to create in-situ freehand AR applications to interact with 3D contents \cite{10.1145/3472749.3474769}.
3D content plays an important role that is augmented for memory reconstruction by immersive technologies in prior works. 
For example, Li et al. demonstrated how AR can be used to support intergenerational memory storytelling by augmenting photos, videos, music, and 3D models that recount past experiences in AR \cite{10.1145/3610903}.
Kang et al. created hybrid tangible AR souvenirs with different input modalities of AR (e.g., hand gestures and voice) that combine a physical firework launcher and AR models, improving the connection between physical souvenirs and their contexts \cite{10.1145/3526114.3558722}.
Considering the advantages of immersive technologies in terms of high-fidelity 3D display and interaction methods aimed to augment 3D contexts, we pioneer the use of AR in reconstructing personal memorable items, by introducing specifically designed interactions to achieve personalized physical interactivity reconstruction for memory archiving.

\subsection{Enriching Physical Interactivity in Virtual Environment}
Physical interactivity in virtual environments allows users to manipulate virtual objects realistically, where the virtual environment responds to the user's behaviors according to physical laws, such as the effects of gravity, friction, inertia, collision dynamics, mechanical structures, and materials \cite{xie2024physgaussian,zhang2024physdreamer,10.1145/3641519.3657448,9879997}. This provides users with more realistic and intuitive experiences in virtual environments.
Ideally, achieving the most realistic physical interactivity requires the combined use of a powerful physics engine and fine-grained modeling of 3D assets.

Prior works have explored automatic approaches through ML models to infer 3D assets' materials to model their physical dynamics. 
These methods utilize visual data from the real world to identify materials and incorporate material-aware dynamics, enabling physically plausible interactions for 3D assets.
For example, \textit{PhysGaussian} seamlessly integrates physically grounded Newtonian dynamics within 3D Gaussians to achieve high-quality motion synthesis \cite{xie2024physgaussian}.
\textit{PhysDreamer} is a physics-based approach that endows static 3D objects with interactive dynamics by leveraging the object dynamics priors learned by video generation models \cite{zhang2024physdreamer}.
\textit{VR-GS} enables real-time execution with realistic dynamic responses on virtual objects by developing a physical dynamics-aware interactive Gaussian Splatting in a VR setting \cite{10.1145/3641519.3657448}.
These methods require precise specification of boundary conditions or material properties of the object to be simulated as a pre-requirement in addition to the visual input. 
These boundary conditions are essentially locations and magnitudes of forces, and fixed locations that specify how the interaction will proceed \cite{zhang2024physdreamer,
xie2024physgaussian,li2023pacnerfphysicsaugmentedcontinuum}. 
Material properties, such as Young's Modulus and Poisson ratio, are also specified.
Combining this information allows the system to predict how the object will respond to forces, i.e., stretch, compress, or even fracture. 
These specifications are traditionally designed by mechanical engineers, based on the model's use case and requirements of the simulation \cite{xie2024physgaussian,zhang2024physdreamer}. 
We cannot expect end-users to have this specialized knowledge, and thus this is outside the scope of our work.
If there are discrepancies between the simulated interactions and the physical world, it remains difficult for end-users to make effective edits.
Processing more complex objects (e.g., objects with mechanical structures, objects with non-uniform materials) often requires skilled modelers or designers to manually rig models using professional software (e.g., Blender, Maya, etc.), and to integrate physical properties using physical engines embedded within game engines (e.g., Unity3D and Unreal Engine) \cite{millington2010game}.

Considering that our work is aimed at end-users without specialized knowledge and addresses the need for personalized 3D physical interaction reconstruction, our prototype first introduces a combined manual and automated 3D assets segmentation method \cite{10.1145/3586183.3606723}. Then we employ predefined physical constraints in the AR interface to simplify and abstract the physics engine, making it easier and accessible for users without domain expertise to create physical interactions based on their own understanding of the objects. \revision{As most related to our work in terms of AR functions, \textit{GesturAR} enables users to create interactive hand gestures for virtual objects \cite{10.1145/3472749.3474769}, while \textit{Ubi Edge} allows end-users to customize edges on daily objects as TUI inputs to control varied digital functions \cite{10.1145/3544548.3580704}. Different from their explorations on specific interactive functions, our prototype aims to achieve a more comprehensive pipeline to allow users to reconstruct and customize physical objects in AR in terms of geometry, interface, and embedded content. Our prototype, with its emphasis on realistic reconstruction within the virtual environment, also provides a comprehensive description of future user-defined interactive virtual objects, originating from the physical world.}

    \section{Identifying Key Interactive Attributes for Memorable Items}
We first conducted a formative study to identify the key physical interactivity attributes of memorable items with the aim of reconstructing interactive digital replications.



\subsection{Participants and Procedure}


Ten participants were recruited from a local campus (5 male, 5 female; age: avg = 25.80, std = 4.83). 
We conducted a semi-structured interview to investigate participants' expectations on the key interactivity attributes of memorable personal items they wanted to preserve in reconstruction.
Each interview session took about one hour and each participant received financial compensation based on local standards. 

After completing the consent form and demographic questionnaire, participants were asked questions on 1) the metadata (names, creation or ownership dates, stories behind, types, etc.) of the physical memorable personal items brought by them, 2) key factors, especially related to the real-world interactivity, that make these memorable personal items meaningful to participants' memories, 3) the detailed expectations when converting the memorable personal items to digital replications, and 4) the potential usage scenarios of the digital replications.
The participants were also asked to use pencils and papers provided by experimenters to sketch their ideas, especially for the detailed expectations when converting the memory artifact to a digital replication.
The experimenter took field notes and video-recorded the whole interview.

\subsection{Results}
We collected 1) video recordings of the interviews (10 hours in total), 2) field notes taken by the experimenter, and 3) sketches created in the interviews from all participants.
The video recordings were transcribed to paragraphs using a commercial ASR system (iFLYTEK6\footnote{iFLYTEK6: https://www.iflyrec.com/zhuanwenzi.html}) and checked by the research team for correctness. 
The data were analyzed using the reflexive thematic analysis method \cite{braun2006using}. Below we summarized the interview results.

\subsubsection{R1: Two prominent types of memorable personal items}
Two types of memorable personal items were frequently mentioned by participants (N=10), which are physical artifacts (non-electronic devices) (e.g., diaries, albums, toys, boxes to keep physical photos, cards, ornaments, artworks, etc.) and electronic devices (e.g., music players, cassette/record players, game controllers, game consoles, instant cameras, film cameras, etc.). 
Both types of memorable personal items were mentioned to require a significant need to reconstruct their physical interactivity features.
For example, P6 mentioned, \engquote{I cherish my old MP3 player, which was my companion for several years. Unfortunately, it's now broken. I would like to recreate a digital copy that can also play the music from my damaged MP3.}
Our participants emphasized that these items hold value not only because they evoke memories when seen but also because they were used or interacted with regularly in the past.
When these items break or become obsolete, users may lose the majority of their traces in memories.  





\subsubsection{R2: Physical traces and transformations for memorable personal items}
The usage traces found on memorable personal items revealed the past interactions between the artifact and the owner. 
The personal traces also represent the ownership of the memory artifact and distinguish the objects with personal memories from brand-new ones. 
Most participants (N=9) mentioned that scratches and traces of wear and tear on mementos can swiftly trigger associated memories. 
For instance, a scratch might symbolize \engquote{a past misuse} (P1), and the traces of wear and tear could indicate a frequently used part of the object. 

For physical artifacts, transformations triggered by motions also contribute to making them more distinct and memorable, such as \engquote{adjusting a Transformer's arms to change its pose} (P8), \engquote{loading a toy gun} (P2), or \engquote{touching the wind chime to make it sway} (P9), and so on.
When discussing digital replicas, participants wanted these copies to maintain the ability to undergo transformations through gestures or motions, emphasizing that \engquote{digital versions should preserve physical interaction like their real counterparts instead of being static displays in museums.} (P9). 



\subsubsection{R3: Distinct Interfaces of memorable personal items}
Special and vintage interfaces of electronic devices (e.g., music players, cassette/ record players, game controllers, game consoles, instant cameras, film cameras, etc.) were also mentioned to enhance the concreteness and tangibility of personal memories.
Participants tend to reminisce about the distinct ways of engagement and the interface design of vintage electronic devices, together with their interactions with the tangible widgets like \engquote{gaming console directional pads} (P10), \engquote{vintage record player sliders and knobs} (P1), and \engquote{distinctive music player buttons} (P3). 
However, with the advent of more advanced technologies like smartphones, these physical widgets and their interface designs may become obsolete. 
P3 observed that \engquote{encountering older TVs, which allow channel switching through a rotatable knob, has become increasingly rare.}
Furthermore, these widgets are often triggered by their internal mechanical structures prone to damage, \engquote{making them easily breakable} (P2, P3, P8). 
For these reasons, the participants stressed the crucial importance of preserving the unique interfaces of electronic devices. 

\subsubsection{R4: Embedded content of memorable personal items}
The embedded content of memorable personal items was viewed as cherished assets by participants and needed digital preservation. 
For artifacts of electronic devices, songs in music players, video games and software in-game consoles, and old photos in a film camera, were mentioned by almost all participants to be the important part of their memory. 
However, the continuous iteration and updates of devices have made it challenging to \engquote{access these older contents on newer devices.} (P1). 
If the old devices were accidentally damaged, \engquote{these contents might become permanently inaccessible. } (P6). 
For physical artifacts, the embedded content included contextual details that evoke and symbolize significant places, times, things, people, and experiences, such as \engquote{the game mechanics of card games} (P6), \engquote{the rules of using a Kendo sword \footnote{Kendo: https://en.wikipedia.org/wiki/Kendo}} (P2) and \engquote{the usage scenario of a wooden table} (P3). 
For example, P3 mentioned, \engquote{Every time I see the table, I'm reminded of the experiences from my first year at university.}. 
Seven participants expressed the desire to incorporate the embedded content into the mementos' digital copies.
This preference arises from the vulnerability of such content to loss and its often implicit nature (e.g., the contextual details of an ancient sword), which poses challenges to its long-term preservation and accessibility.

\subsection{Summary}
The formative study uncovers participants' expectations regarding the key physical interactivity attributes of memorable personal items when creating their digital counterparts.
\textit{Physical artifacts} and \textit{electronic devices} emerge as two significant types of memorable personal items that are instrumental in evoking personal memories.
Participants noted that the \textit{physical traces and transformations} symbolized the memorable personal items' uniqueness, linking them closely to personal memories. 
Preserving the unique and vintage \textit{interfaces of older electronic devices} is essential for reinforcing the memories' concreteness and tangibility. 
Furthermore, the \textit{embedded content} is highly valued by participants and crucial for digital reconstruction.
Overall, the findings suggest the need to explore user-oriented and memory-evoked digital replications of memorable personal items, emphasizing their physical interactivity.

    \section{Designing and Prototyping Interactive Digital Items}

\revision{Motivated by the findings of our formative study, we recognize the key attributes of interactivity that should be integrated into digital reconstructions of memorable personal items.}
Thus, we define and elaborate on the concept of \textbf{I}nteractive \textbf{D}igital \textbf{I}tem (IDI) as digital reconstructions with physical interactivity features. 
We proceed to outline the expected features of IDI and introduce \emph{InteRecon}, a user-oriented prototype designed for creating IDI.
InteRecon is designed to cater to individual end-users who have been identified as the most significant users of personal memory archives and extensions. 
Target items for reconstruction should be cherished physical mementos capable of evoking personal memories \cite{kirk2010human,10.1145/3173574.3173998,10.1145/3544549.3585588}. 
IDI emphasizes digitally inheriting the physical interactivity from its physical counterparts to enhance the efficacy of digital reconstructions in presenting personal memories.

\subsection{Concepts and Design Goals}
Based on our formative study, we have formulated three \textbf{D}esign \textbf{G}oals to delineate the physical interactivity of IDI to be reconstructed in the digital realm.


\subsubsection{\textbf{DG1: Reconstructing Geometry}}
IDI should reconstruct geometric properties to accurately mirror the entire physical appearance and transformations of its real-world counterparts (R2). 
In this case, we propose that IDI should feature a 3D model faithfully replicating the entire physical appearance (R2) while also emulating the physical properties, such as gravity and collisions. 
Furthermore, for complex physical artifacts connected by joints (R1), IDI should enable interaction through natural hand movements (e.g., pushing, pulling, manipulating finger joints, etc.) to create realistic motion transformations, including movement and rotation, aligning with the item’s joint mechanism.



\subsubsection{\textbf{DG2: Reconstructing Interface}}
For electronic devices, IDI should reconstruct the interfaces of its physical counterpart (R3), including the tangible widgets along with the internal programs (e.g., the functions or effects triggered by certain physical operations) on these devices. 
When users interact with these virtual tangible widgets (e.g., pressing buttons, dragging sliders, etc.), IDI should replicate analogous effects in the virtual interface (e.g., switching to the next song in an MP3 player, moving avatars in a gaming console, etc.) mirroring real-world interactions.

\subsubsection{\textbf{DG3: Reconstructing Embedded Content}}
IDI should also preserve the content stored or embedded within the item.
For electronic devices, digital content, such as songs, photos, movies, and games, stored in the devices need to be reconstructed. 
For physical items, the embedded content could be the contextual information (e.g., usage scenarios, associated individuals, and related stories, etc.) in the form of notes, photos, or videos.
Users should be able to access and associate this content with IDI.
For example, consider a bicycle with a scratched frame; in such cases, users can annotate the scratch with a brief note describing the incident that caused the scratch, and users can also give general comments on the bicycle (e.g., the history, interesting events, etc.). 
As validated by our formative study (R4), reproducing the content and making it accessible in IDI could be instrumental for both functional integrity and memory preservation.

\subsection{InteRecon: An Prototypical Application for IDI Creation}
\label{design_interRecon}
We designed InteRecon, an AR application for end-users to create IDI for the interactivity-aware reconstruction of memorable personal items. InteRecon features four functions corresponding to the above design goals.


\subsubsection{Function 1: Reconstructing 3D Appearance}
\label{sec: Reconstruct the 3D model by 3D scanning}
\begin{figure}[tbh!]
     \centering
     \includegraphics[width=\linewidth]{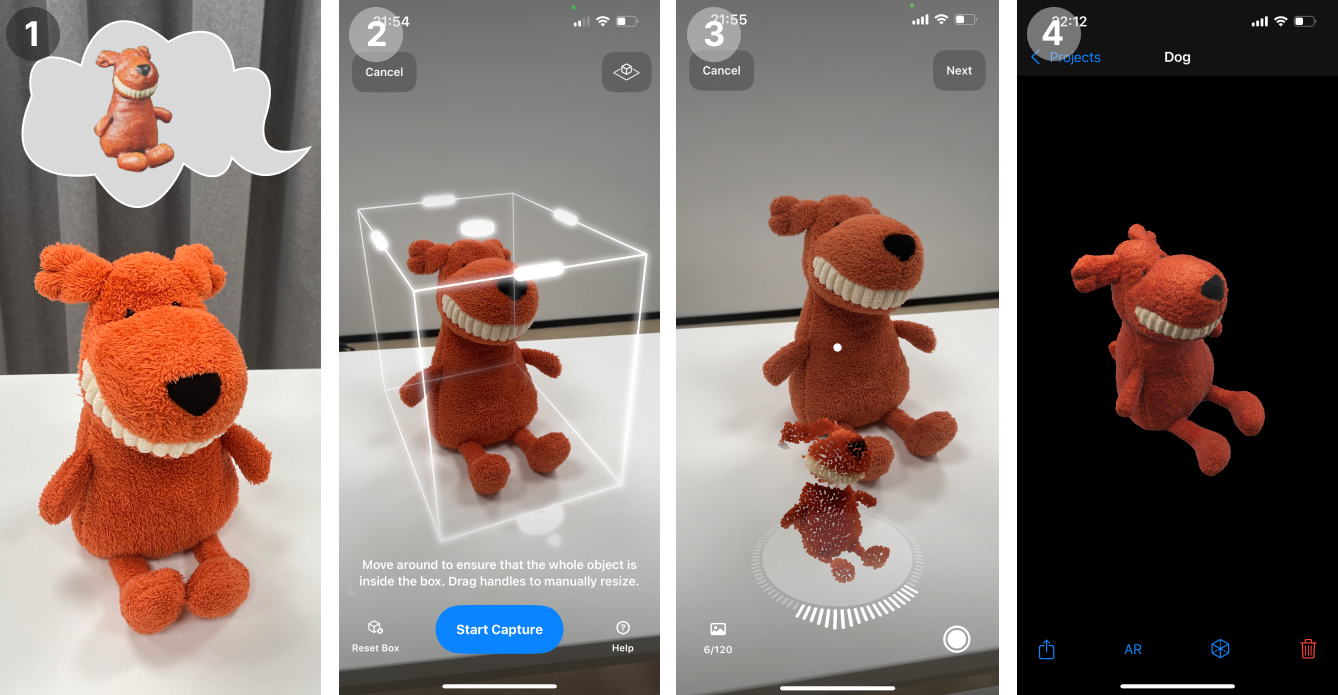}
     \vspace{-2ex}
\caption{\textbf{The interactive process for reconstructing 3D appearance. (1) The user opens the application and points it at the target object. (2) An automatic bounding box is generated around the target object in the application interface. (3) Circle around the object with visual guidance on regions to capture more images. (4) A 3D reconstructed model is ready for further use.}}
\Description{Figure 2 has 4 sub-figures (a, b, c, d, from left to right) in a row, showing the interactive process for reconstructing the 3D model. (1) shows a user opens the mobile application and points it at the target object. (2) shows the screen of the mobile application, an automatic bounding box that is generated around the target object in the application interface. (3) shows the screen of the mobile application when scanning objects, a circle around the object with visual guidance on regions where the algorithm needs more IDIs. (4) shows a 3D reconstructed model which is ready for further use.}
\label{fig:3dapp}
\end{figure}

To support faithfully capturing the physical appearances for IDI (\textbf{DG1}), we developed a mobile application to enable users to reconstruct the 3D model of the target object through 3D scanning.
The Fig. \ref{fig:3dapp} illustrated the entire scanning process. 
First, the user needs to open the application and point it at the target object. 
An automatic bounding box with the length, width, and height of the object is generated before capturing. 
Then the user can move the mobile phone slowly to circle around the object while the application automatically captures the right image for reconstruction. 
The app provided visual guidance on regions where the algorithm needs more images, along with additional feedback messages to help the user capture the best quality shots. 
After finishing one orbit, the user can flip the object to capture the bottom. 
Once scanning for three orbits (front, side, and bottom surface of the bounding box) is completed, the application will proceed to the reconstruction stage, which runs locally on the mobile device. 
A 3D reconstructed model will be ready for further use. 

\begin{figure*}[tbh!]
\centering
\includegraphics[width=\textwidth]{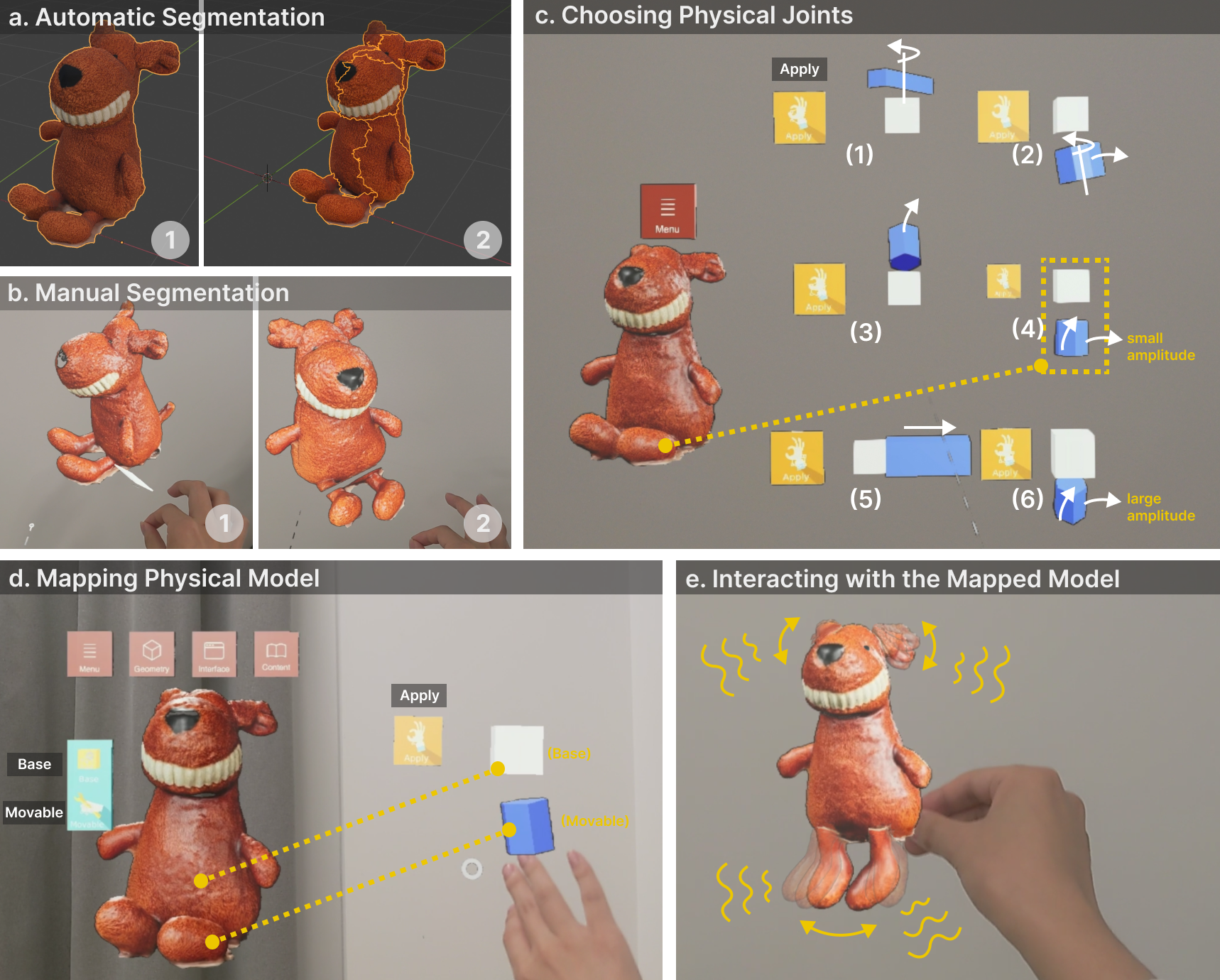}
\caption{\textbf{The interactive process of adding physical transforms. (a-1,2) Segment the puppy's model using an automatic approach. (b-1,2) Segment the puppy model in AR by using a segmenting plane and breaking down the model's leg along the plane. (c) Use hands to touch the blue cube to observe the movement features of each demonstrated joint until identify one that resembles the model's leg. Detailed introductions to each joint are listed in Table \ref{tab:joints}, and the numbers close to each joint in the figure correspond to the numbers in Table \ref{tab:joints}. (d) Select the `movable' segment of the model (the leg) using a pinching gesture and confirm the choice by clicking the `movable' button, which is analogous when selecting the `base' (the body). Press the `Apply' button to apply the joint to the model's leg and body. (e) After repeating the above processes for mapping the model's ear joints, shake the model's body and generate similar motions to a real-world toy puppy's legs and ears in AR.}}
\Description{Figure 3 illustrates The interactive process for adding physical transforms. (a-1,2) Segment the puppy model's leg in the software. (b-1,2) Segment the puppy model in AR by using a segmenting plane and breaking down the model's leg along the plane. (c) Use hands to touch the blue cube to observe the movement features of each demonstrated joint until identify one that resembles the model's leg. Detailed introductions to each joint are listed in Table \ref{tab:joints}, and the numbers close to each joint in the figure correspond to the numbers in Table \ref{tab:joints}.
(d) Select the 'movable' segment of the model (the leg) using a pinching gesture and confirm the choice by clicking the 'movable' button, which is analogous when selecting the `base' (the body). Press the `Apply' button to apply the joint to the model's leg and body.
(e) After repeating the above processes for mapping the model's ear joints, shake the model's body and generate similar motions to a real-world toy puppy's legs and ears in AR.}
\label{fig:geo}
\end{figure*}

\begin{figure*}[tbh!]
\centering
\includegraphics[width=\linewidth]{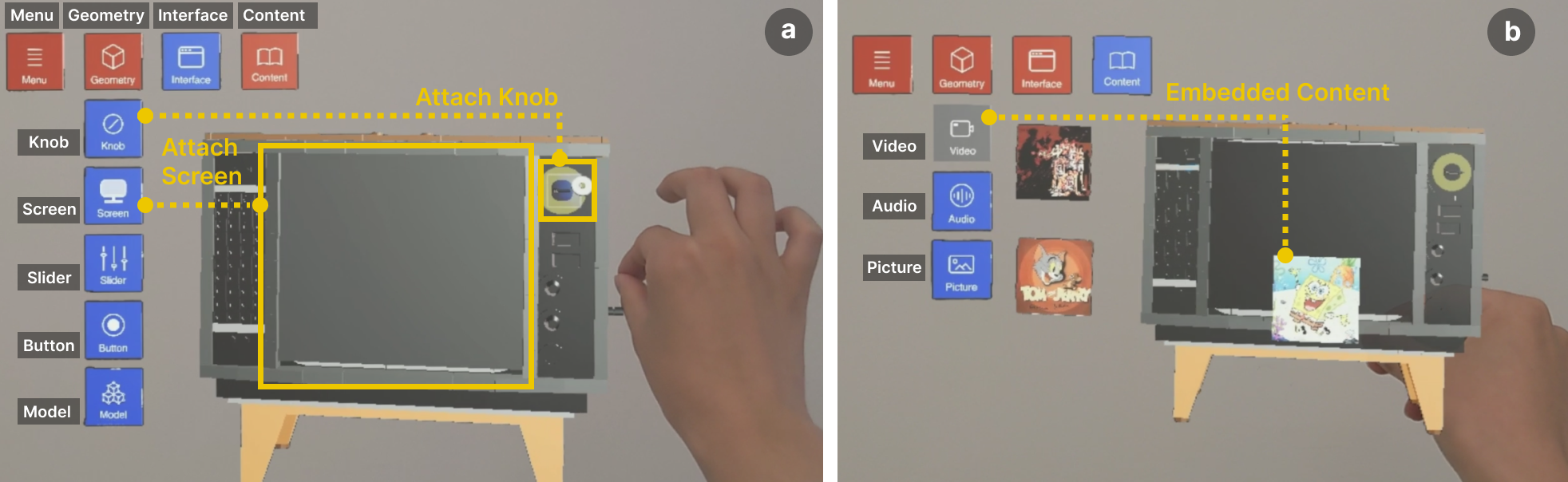}
\caption{\textbf{The AR interface for the functions of reconstructing the interface and adding embedded content.  (a) The `interface' in the menu includes four commonly used widget categories: `knob', `screen', `slider', and `button', and a target model category that the user scanned by Function 1 (in this case, the TV model). Press the `knob' and `screen' buttons to select a knob widget and a screen widget and attach them to the TV model to reconstruct its interface. (b) The `content' includes three categories in the menu: `video', `audio', and `picture'. 
Press the `video' button to release the three embedded videos that users uploaded and drag them to the TV model to import. }}
\Description{Figure 4 illustrates the AR interface for the functions of reconstructing the interface and adding embedded content.  (a) The `interface' in the menu includes four pre-defined widget categories: `knob', `screen', `slider', and `button', and a target model category that the user scanned by Function 1 (in this case, the TV model). Press the `knob' and `screen' buttons to select a knob widget and a screen widget and attach them to the TV model to reconstruct its interface. (b) The `content' includes three categories in the menu: `video', `audio', and `picture'. 
Press the `video' button to release the three embedded videos that users uploaded and drag them to the TV model to import. }
\label{fig:inter}
\end{figure*}

\subsubsection{Function 2: Adding Physical Transforms}
As proposed by \textbf{DG1}, a further step to realize interactivity reconstruction should focus on the modeling physical dynamics and mechanical properties despite static appearances. 
This allows users to segment the mesh reconstructed from the first function and apply physical constraints to these segments, simulating the object's physical dynamics and mechanical properties. Under these constraints, when the model is directly touched by hands, its segments will respond just as they would in the real world.
A two-stage pipeline is devised to achieve this function, as illustrated in Fig. \ref{fig:geo}. 
First, the user should segment the model to get it prepared to apply the physical constraints on it.
To achieve mesh segmentation, a segmenting plane is provided by InteRecon in AR to support the user in breaking down the mesh model along the plane, as is shown in Fig. \ref{fig:geo}(b). 
The user can use their hands to adjust the size, the rotation, and the location of the plane to better match the potential segment position on the mesh. Alternatively, we also incorporated an unsupervised automatic segmentation approach, inspired by Style2Fab~\cite{10.1145/3586183.3606723}. This segmentation method is based on spectral segmentation, which leverages the mesh geometry to predict a mesh-specific number of segments. Users can choose to either segment by providing the planes in AR, or use the automatic segmentation method.

After acquiring a segmented model in AR, the user could move on to the second stage to further map the physical joints to IDI to simulate the mechanical transforms on the item.
To help the user rapidly map the joints, pre-defined mechanical structures of joints \cite{blake1985design} were implemented in the application, as is shown in the Appendix Table \ref{tab:joints} and their virtual representation in Fig. \ref{fig:geo}(c).
Each of the virtual joints includes two cubes, indicating the relative movement in one degree of freedom and restricting movement in one or more others. 
We also offer different resistance options for each type of joint to accommodate various situations. For instance, although the movement mechanism of the leg joints in soft toys and LEGO figures is similar, the resistance in LEGO figures is greater, which means that under the same amount of force, their range of motion would differ. In order to illustrate relative movements more effectively, the two cubes are distinguished by different colors, with the grey cube representing the `base' cube (non-movable) and the blue cube representing the `movable' cube (capable of motion).
The user could use their hands to touch the blue cube to observe the relative movement features of each demonstrated joint until they identify one that resembles the physical joint.
The user could select the `movable' segment of the mesh model using a pinching gesture and confirm their choice by clicking the `movable' button, as is shown in Fig. \ref{fig:geo}(d).
This process is analogous when selecting the `base'.
To conclude the process, the user can click the `Apply' button next to the identified virtual joint to map it onto the mesh model.
InteRecon also supports the incorporation of multiple joints by repeating the aforementioned process.

\subsubsection{Function 3: Reconstructing Interface}
InteRecon enables users to reconstruct the interfaces of electronic devices on IDI by integrating widgets to the 3D model (\textbf{DG2}). 
We illustrated the interactive process of this function in Fig. \ref{fig:inter}(a). 
The user can apply \textbf{Fuction 2} on the model of an electronic device to segment the tangible widgets from the main body of the model while defining their mechanical transforms (e.g., rotations for knobs) to simulate real-world effects of the widgets. 
Further, the user can edit widget models and corresponding analogous effects to simulate certain interface logic. They first press the `interface' button to view widget categories and then select widgets from four pre-defined categories: `knob', `screen', `slider', and `button'. 
Each category contains widget instances corresponding to different functions in different device models. 
For example, the `screen' category includes two types: one is a display that can present content such as photos and videos; the other is a camera's viewfinder, allowing you to see what's being photographed.
The user needs to refer to the widgets on the physical electronic device and spawn similar virtual widgets (with the same functionalities compared to the actual buttons) in AR by clicking. 
Afterward, the user can move the position and adjust the size of the virtual widgets to make sure they have a similar relative position and size to their physical counterparts.
The `attach' button allows users to bind virtual widgets generated by InteRecon to the model of the physical widgets.
After binding, the widget model could control the content uploaded from \textbf{Function 4}.
The `invisible' button could make the virtual widgets invisible while preserving their functions that could be triggered in their positions on the surface of the 3D model in AR. 
Thus, the user could trigger the function on the surface of the 3D model without seeing the virtual widgets.

\subsubsection{Function 4: Adding Embedded Content} 
We devise a content management system to help the user better manage the content associated with their memorable items. 
The user can upload different forms of content, such as songs, videos, pictures, and texts.
Then these contents are synchronized to the `Content' category in the AR environment by our researcher.
The user could select the target content in AR and move it to a specific IDI model to incorporate the embedded content of the item.
Fig. \ref{fig:inter} (b) shows the AR interface of adding embedded content to an IDI model.

\subsection{Implementation}
\label{implementation}
We built InteRecon's mobile applications on iPhone 14 Pro to implement InteRecon's functions (\emph{Function 1} and \emph{Function 4}) on mobile devices. For 3D object scanning, we utilized the reality kit (Object Capture API~\footnote{RealityKit: https://developer.apple.com/documentation/realitykit}) from Apple Developer to build an app that runs on iOS 17.0+. 
For the AR interface of InteRecon, we implemented on HoloLens 2, which was connected to a PC (with an Intel i9-12900K CPU and an RTX A6000 24G GPU) using a wireless network using Unreal Engine Version 4.26 \revision{to handle complex geometric meshes}.
We integrated the Mixed Reality Tool Kit (MRTK~\footnote{MRTK: https://github.com/microsoft/MixedReality-UXTools-Unreal}) to handle the hand interaction and UI elements in the application. 

We used UE's built-in physical engine to implement the physical effects such as gravity, collision, physical joints, and hand manipulations, of IDI. 
Specifically, for the manual mesh segmentation, we first converted the scanned mesh to a \emph{Procedural Mesh} and called the \emph{Slice Procedural Mesh} method to cut the mesh with a hand-held cut plane at runtime. 
Segmented meshes were stored both as an array of \emph{Procedural Mesh} in the UE program with further physical operations enabled and as static mesh copies in the disk. 
For automatic segmentation, we use the approach from Style2Fab \cite{10.1145/3586183.3606723}, which uses a spectral decomposition approach. It leverages the spectral properties of a graph representation of the 3D mesh to identify meaningful segments. \revision{By analyzing the eigenvalues and eigenvectors of the graph's normalized Laplacian matrix}, this technique uncovers the mesh's inherent structure, grouping similar vertices together to create a meaningful segmentation of the model. \revision{"Meaningful segments" refers to portions of the mesh that are grouped together based on their similarity or shared properties in geometric, semantic, structural, or functional purpose. For example, in the case of a 3D model of a human body, the model will be divided into segments like limbs and a torso. For an articulated structure, the object will be divided into different components that serve specific functional purposes, such as an elastic linkage part and a non-functional decoration part.}  This method doesn’t require training and can be generalized across a wide range of 3D models.
Joint creation is enabled using the built-in physics engine of the Unreal Engine. 
The hand interaction related to physical effects, such as a slight touch on objects, is implemented by binding a collision sphere to the tip of the finger in AR. 

    \section{User Study}
We conducted a two-session user study to understand the feasibility of using InteRecon to create IDI and further explore the participants' approaches to using InteRecon to reconstruct their own items, collecting feedback on the challenges, future opportunities, and applications of IDI. 

In the first and second sessions, we invited 16 participants from a local university campus (8 male, 8 female; age: avg = 24.13, std = 2.28). 
\revision{In the appending study, we recruited 10 participants through questionnaires from the campus, aiming for a diverse mix of professional backgrounds (4 male, 6 female; age: avg = 24.6, std = 2.5). The participants included 2 VR developers, 2 architects, 1 fashion designer, 1 product manager, 1 industrial designer, 1 hardware engineer, 1 curator, and 1 professor. Notably, three of these participants were re-invited from the previous two sessions.}
All had prior experience with using AR and VR devices with avg = 4.98, std = 1.53 on a scale from 1 (not at all familiar) to 7 (extremely familiar). 
The first two sessions took around 2 hours and the additional brainstorming took around 1 hour.
Each participant was paid an equivalent of 50 USD in local currency for compensation. 
The hardware configurations and AR deployment employed in the study were consistent with those detailed in Sec. \ref{implementation}.
The study was conducted in our laboratory and received ethical approval from the university's ethics review board.


\begin{table}[tbh!]
\centering
\caption{\textbf{Descriptions for the atomic interactions of each function in the session one.}}
\Description{Table 2 shows descriptions of the atomic interactions of each function in the first session.}
~\label{tab:atomic_interaction}
    \vspace{-0.3cm}
    \small
    \resizebox{\linewidth}{!}{
    \begin{tabular}{l|l|l}
    \hline
        \textbf{ID} & \textbf{Function} & \textbf{Interaction Steps} \\ \hline
        A & Reconstructing 3D Appearance & \begin{tabular}[c]{@{}l@{}}1. Align the camera with the object and adjust the bounding box.\\ 2. Move the camera around the object. \\ 3. Examine and confirm the mesh.\end{tabular} \\ \hline
        B & Adding Physical Transforms & \begin{tabular}[c]{@{}l@{}}1. Segment the model for making joints.\\ 2. Touch the pre-designed joints and find a similar joint.
\\ 3. Apply the joints by mapping \textit{Base} and \textit{Movable} cubes.\end{tabular} \\ \hline
        C & Reconstructing Interface & \begin{tabular}[c]{@{}l@{}}1. Select the category of the virtual widget.\\ 2. Attach the virtual widgets on the model. \\ 3. Adjust the size and the visibility of the widgets. \end{tabular} \\ \hline
        D & Adding Embedded Content & \begin{tabular}[c]{@{}l@{}}1. Upload /edit the content with the application. \\ 2. Import the content to the model.\end{tabular} \\ \hline
    \end{tabular}
    }
\end{table}

\begin{table}[tbh!]
    \centering
    \caption{\textbf{Task scripts in the session one. }
    We employed a combination of ID letters and sequence numbers from Table  \ref{tab:atomic_interaction} to reference specific atomic interactions (e.g., the \textit{A1} corresponds to the first atomic interaction in the `Reconstructing 3D Appearance' category).}
    ~\label{tab:scripts}
    \Description{Table 2 shows the task scripts in the first session. We employed a combination of ID letters and sequence numbers from Table  \ref{tab:atomic_interaction} to reference specific atomic interactions (e.g., the \textit{A1} corresponds to the first atomic interaction in the `Reconstructing 3D Appearance' category).}
     \vspace{-0.3cm}
    \resizebox{\linewidth}{!}{
        \begin{tabular}{l|l|l}
        \hline
        \textbf{ID} & \textbf{Task} & \textbf{Script} \\ \hline
        T1 & Scan a model of a toy Panda & A1-A2-A3 \\ \hline
        T2 & Reconstruct a Moon lamp IDI’s physical transforms & B1-B2-B3 \\ \hline
        T3 & Reconstruct a TV IDI’s interface & C1-C2-C3-(D1-D2)\textasciicircum{}n \\ \hline
        \end{tabular}
    }
\end{table}

\subsection{Session One: Full Functions Experiencing}
\label{session-one}
The goal of this session was to assess the feasibility of utilizing InteRecon to create the IDI within a mixed reality environment. 
To achieve this, we asked participants to experience all the functions provided by InteRecon to inspire them to propose more personal IDIs and usages in the next session.
We broke down InteRecon's functions described in Sec. \ref{design_interRecon} into one-step interactions and designed scripts of one-step atomic interaction sequence for each function, as illustrated in Table \ref{tab:atomic_interaction}. 
We further designed three micro IDI-creation tasks, which collectively cover all the atomic interactions across the four functions.
Each task was structured to progress through a list of several atomic interactions, as outlined in Table \ref{tab:scripts}. 
In the case of T3 from micro-tasks, as there were multiple contents in a TV to be added, we anticipated that the D function in Table \ref{tab:atomic_interaction} would be iterated \textit{n} times for participants to reconstruct a TV IDI.
We noted this in our task scripts in Table \ref{tab:scripts}. 
Additionally, we created tutorials for each task, available in both the AR environment and print format. 

We first asked the participants to walk through the Hololens 2 official tutorial to learn how to navigate the user interface with basic hand gestures. 
After finishing the consent form and demographic questionnaire, participants were first introduced to the study and provided with a tutorial. 
They were then guided to complete three tasks sequentially. 
Before each task, participants were invited to interact with the relevant physical item associated with the task. 
This allowed them to gain an understanding of the item's interactivity, such as the widgets on the TV and the joint mechanism between the Moon lamp's base and body, assisting them in reconstructing these interactions in AR.
After completing each task, participants provided brief feedback on their InteRecon experience and were granted a 10-minute break.
The study involved two experimenters: one was responsible for introducing the study and aiding participants in tutorial comprehension, while the other monitored participants' Hololens views and documented the feedback for each task. 
After finishing all the tasks, the participants were asked to complete a questionnaire with Likert-type (scaled 1-7) questions on evaluating the feasibility of InteRecon's four functional categories.
The questions with metrics employed in the questionnaire are detailed in the Appendix Table \ref{tab:questionnaire}.

\subsection{Session Two: Free Exploration on Prototyping Personalized IDI}
\label{session-two}
The goal of this session was to investigate how participants employ InteRecon to create their own IDI in an exploratory manner without specific tasks, aiming to obtain further feedback and ideas on the challenges, application scenarios, and future opportunities of InteRecon.
We conducted a phone interview to inquire about the types of items they were interested in reconstructing before participants arrived for the user study.
This ensured the InteRecon's resources, the elements within the categories of `geometry', `interface', and `content', could accommodate the potential interactivity features of reconstructing participants' items. 
Three participants said that the items they wished to reconstruct were not at hand, so with the assistance of our researcher, they downloaded similar 3D models from the internet to proceed with the next steps of creating IDI, bypassing the step of scanning the physical item. 
We also provided participants with items mentioned in our formative study in case the participants had some more creative ideas to implement.
Participants were then asked to utilize InteRecon to recreate the IDI.
This exploratory session lasted approximately 30 minutes. 

At this stage, participants were already familiar with InteRecon and an experimenter was present to address any questions they might have had. 
Participants were asked to ``think-aloud'' \cite{van1994think} to express their thoughts promptly during the process. 
In the end, we held in-depth interviews (30-40 minutes) with our participants regarding their qualitative feedback.
The entire session was recorded on video by the experimenter.

\subsection{\revision{Appending Study: Gaining Design and Usage Implications from Broader Audiences}}
\revision{We conducted an appending study to investigate broader designs and applications for IDI, expanding its conceptual development beyond personal memory archiving. 
Given the diverse professional backgrounds of our participants, they were able to brainstorm potential applications and utilities of IDIs in their professional contexts, contributing ideas on new conceptual developments for IDIs in the future.}

\revision{The experimenter introduced the concept of IDI first and showcased examples of IDIs created in previous sessions, providing participants with a comprehensive overview.
Following this introduction, participants were asked to try using AR glasses and engage directly with the created IDIs. 
The introduction and try-on took around 30 minutes for each participant.}

\revision{We then conducted semi-structured interviews with the participants, focusing on the following questions: 1) benefits of digital 3D reconstruction for objects while preserving their physical interactivity, 2) potential application scenarios for IDI of their professional areas, 3) specific uses of IDI in professional or functional settings (e.g., museums, education, healthcare, and travel), and 4)extensions of the concept of IDI beyond personal memory archiving.
We specifically emphasized that the participants should incorporate their professional experiences or expertise when answering questions.
The entire procedure was recorded on video by the researcher.}





\begin{figure}[htbp]
    \centering
    \includegraphics[width=\linewidth]{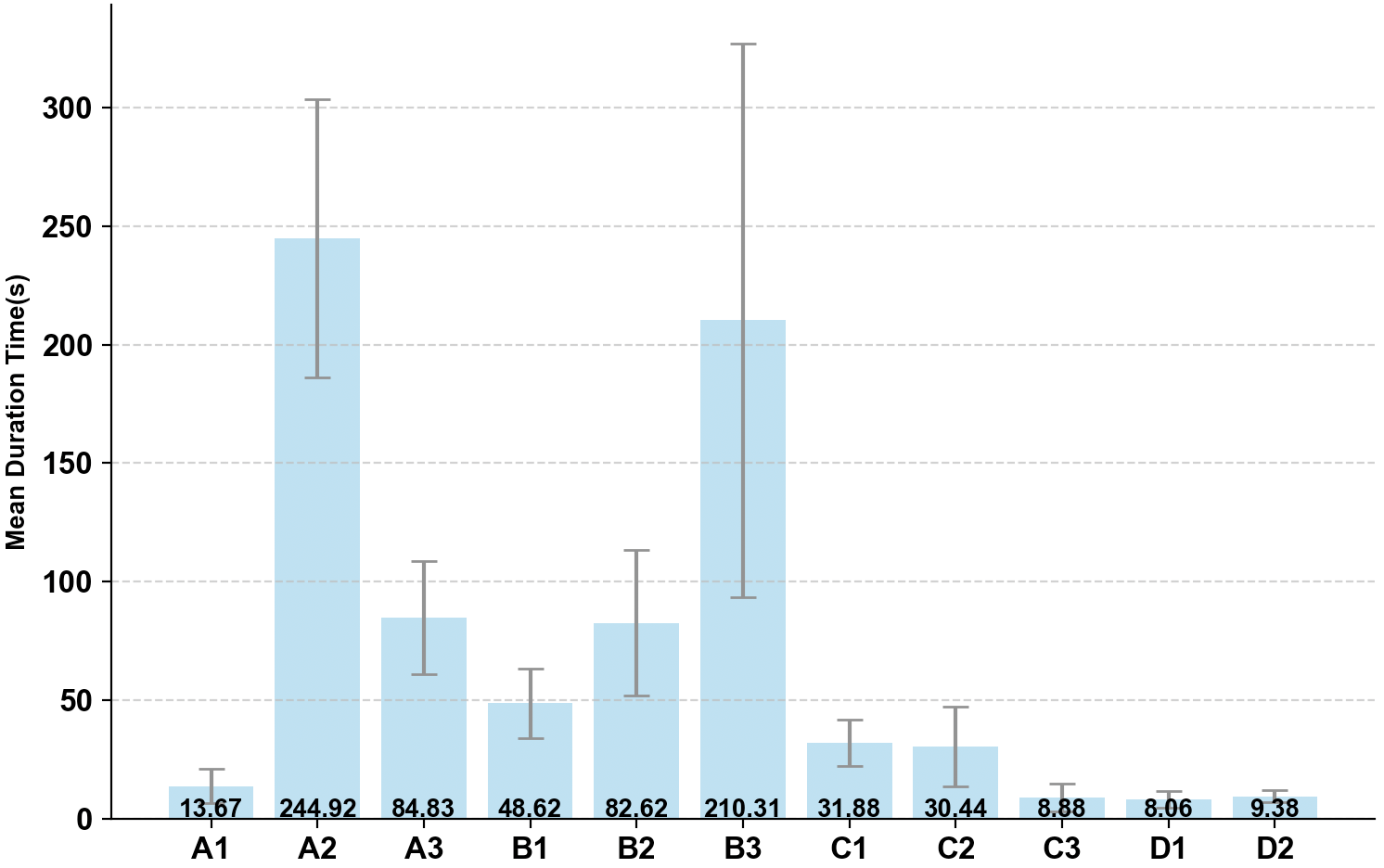}
    \vspace{-2ex}
    \caption{\revision{\textbf{The mean duration time of each atomic interaction in three tasks in Session One.}}}
    \Description{Figure 5 shows the duration of each atomic interaction in three tasks in Session One.}
    \label{duration-time}
\end{figure}

\subsection{Results}
\label{fi:overview_results}
We reported the results collected from our user study, including the duration time of each interaction with failure cases in Session One, the questionnaire consisting of 7-point Likert scale data regarding the user subjective ratings for four categories of InteRecon functions in Table \ref{tab:atomic_interaction}, and the qualitative feedback from the interview in Session Two. 
Fig. \ref{fig:results} illustrates examples of the reconstructed IDIs by our participants.
Except the Moon Lamp in a-1 of Fig. \ref{fig:results} was created in the pre-designed task of Session One (Section \ref{session-one}), the rest examples were created by participants in Session Two (Section \ref{session-two}). 
We conducted a thematic analysis with qualitative feedback data from 16 participants. 
We report the overall results in Sec.~\ref{sec: subj_rating} to assess the feasibility of the authoring workflow of using InteRecon to create IDI and further report the qualitative results in the following subsections.

\subsubsection{\revision{Overall Results}}
\revision{
We recorded the mean duration time to complete the atomic interaction in each task (11 data points in 3 tasks) in Session One, resulting in 176 data points in total across 16 participants, as illustrated in Fig. \ref{duration-time}. 
For the D1 and D2 interactions, which occurred multiple times in the task, we recorded the first trial for each participant.
The results show that the mean duration time of every atomic interaction is within a reasonable time phrase (under 250 seconds).
Notably, A2 and B3 posed a relatively longer duration time for users to complete. 
The difficulty of A2 lies in the fact that the user needs to move the camera slowly to ensure that enough key frames are captured to generate the reconstruction results. Additionally, in order to capture the three orbit data of the object (front, side, and bottom surface), the participant needs to hold the camera and circle around the object three times. 
The difficulty of B3 arises from the need to employ pinching gestures for selecting model segments within the AR environment. 
Participants reported challenges in this selecting process, partly due to lower pinching gesture accuracy.
We also observed 3 failure cases during B3 involving 3 participants and this may be attributed to the confusion in identifying the `Base' and `Movable' segments of the model.
Two failure cases also occurred during the B1 operation in Session Two, as participants attempted to cut overly complex mesh objects in AR (e.g., having a higher number of vertices and intricate topological structures). This complexity led to crashes in the HoloLens system.
}

\label{sec: subj_rating}

\begin{figure}[tbh!]
     \centering
     \includegraphics[width=\linewidth]{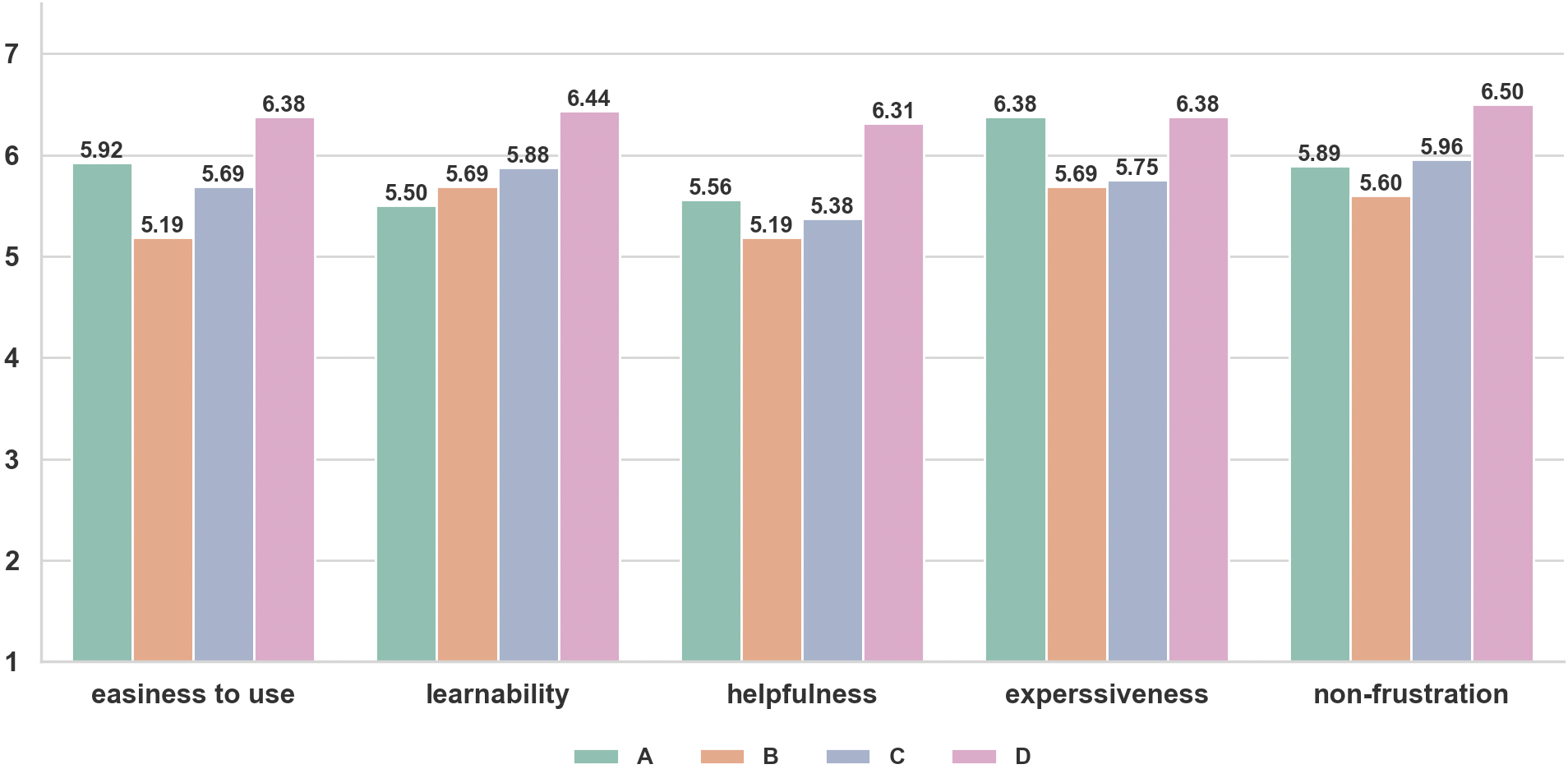}
     \vspace{-2ex}
     \caption{\textbf{Average subjective rating scores for 4 categories of functions in Table \ref{tab:atomic_interaction}. The first (green), second (orange), third (purple), and fourth (pink) columns in each cluster indicate the score distribution across four function categories. 1 - strongly disagree, 7 - strongly agree.}}
     \Description{Figure 6 shows the average subjective rating scores for 4 categories of functions in Table 1. The first (green), second (orange), third (purple), and fourth (pink) columns in each cluster indicate the score distribution across four function categories. 1 - strongly disagree, 7 - strongly agree.}
     \label{BERT}
\end{figure}

\begin{figure*}[tbh!]
\centering
\includegraphics[width=\textwidth]{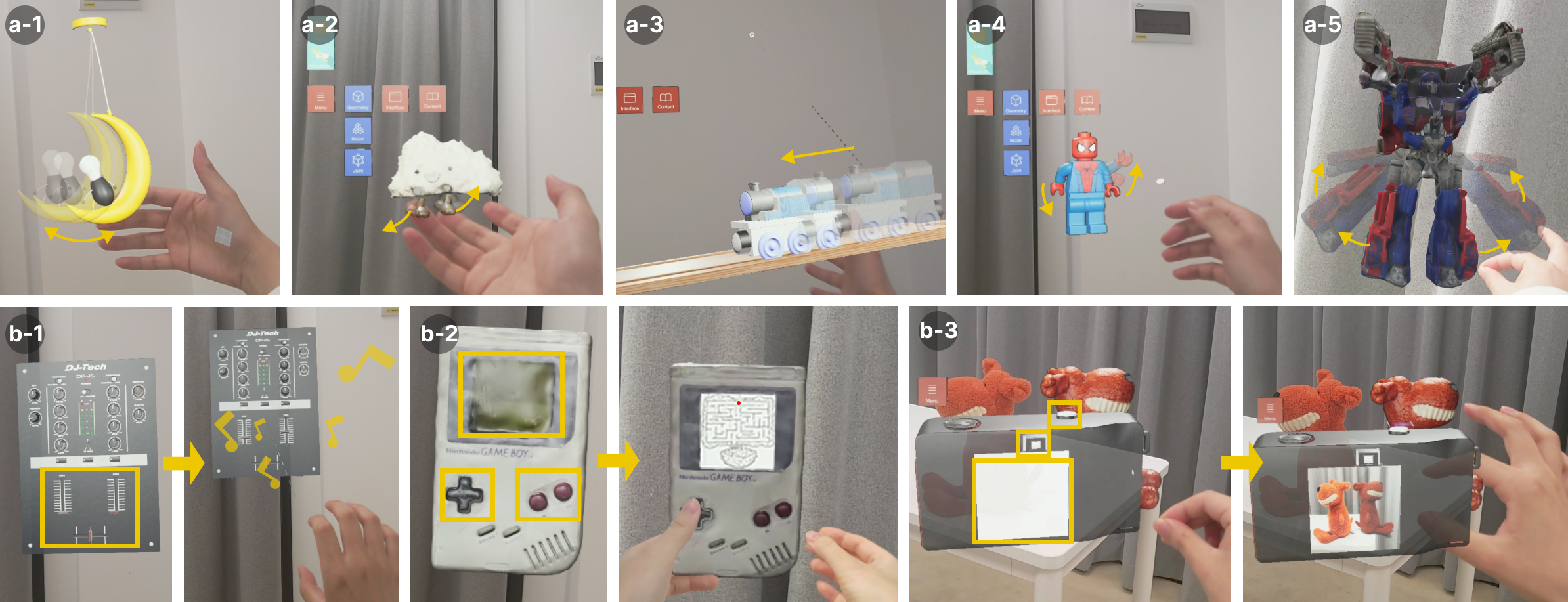}
\caption{\textbf{Example IDIs shown in AR environment created in our user study. (a-1,2,3,4,5) IDIs of physical artifacts including reconstructed physical joints, which could be interacted by hands and generate similar movements to the real world. (b-1,2,3) IDIs of electronic devices, including reconstructed interface with widgets. (b-1) Reconstruct the slider for the DJ booth's IDI: Drag the slider to adjust its volume. (b-2) Reconstruct the screen, the buttons of on/off and directional pads for the Game Boy's IDI: Press the buttons to play the puzzle game. (b-3) Reconstruct the display screen, the viewfinder, and the shutter for the camera's IDI: By pressing the shutter button, a photo of the scene within the viewfinder is captured and displayed on the screen.}}
\Description{Figure 6 shows example IDIs shown in AR environment created in our user study. (a-1,2,3,4,5) IDIs of physical artifacts including reconstructed physical joints, which could be interacted by hands and generate similar movements to the real world. (b-1,2,3) IDIs of electronic devices, including reconstructed interface with widgets. (b-1) Reconstruct the slider for the DJ booth's IDI: Drag the slider to adjust its volume. (b-2) Reconstruct the screen, the buttons of on/off and directional pads for the Game Boy's IDI: Press the buttons to play the puzzle game. (b-3) Reconstruct the display screen, the viewfinder, and the shutter for the camera's IDI: By pressing the shutter button, a photo of the scene within the viewfinder is captured and displayed on the screen.}
\label{fig:results}
\end{figure*}

Participants were asked to provide their ratings towards the four categories of functions through our questionnaire, as illustrated in Fig. \ref{BERT}. 
We employed five metrics: ease of use, learnability, helpfulness, expressiveness, and non-frustration. 
A detailed description of each metric's question can be found in the Appendix Table \ref{tab:questionnaire}. 
Participants found the interaction process user-friendly, as reflected in the high average rating for ease of use (avg = 5.81, std = 0.50). 
They expressed confidence in their ability to use the functions, indicated by a high learnability score (avg = 5.88, std = 0.41). 
The scores of `helpfulness' was also well received (avg = 5.55, std = 0.52). 
Moreover, participants appreciated the expressiveness of the InteRecon, giving it a high score (avg = 6.08, std = 0.42), and they experienced minimal frustration, as shown by the rating for non-frustration (avg = 5.94, std = 0.48). 
These results show that InteRecon is effective, expressive, and user-friendly in creating IDI, with its rich customization possibilities receiving positive feedback from participants.

\begin{figure}[tbh!]
     \centering
     \includegraphics[width=\linewidth]{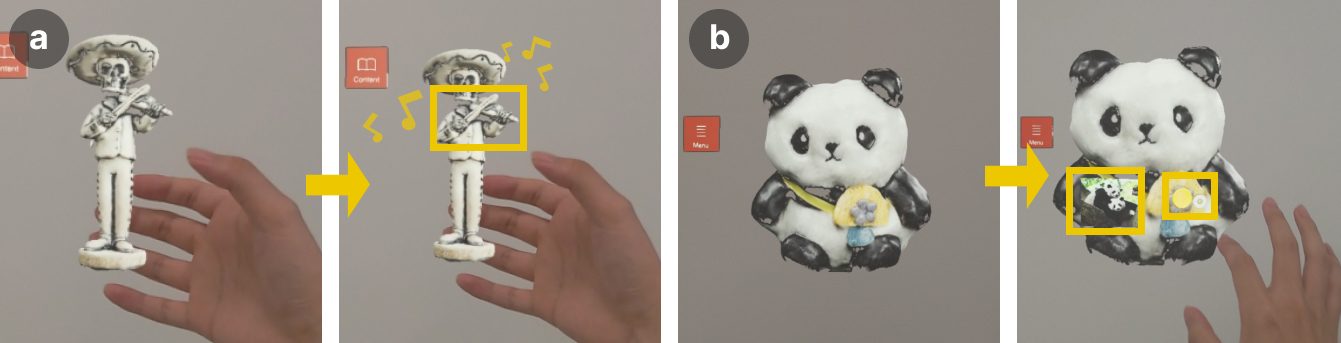}
     \vspace{-2ex}
\caption{\textbf{Two example IDIs participants created that augmented with additional interactivity beyond the real world. (a) `Interactive Statue': the reconstructed violinist statue augmented with its beyond real-world functions of playing music by adding a ‘Play’ button and embedded content. (b) `Interactive Photo Album': the reconstructed souvenir augmented with its beyond real-world functions of displaying photos by adding a button widget and a screen widget.}}
\Description{Figure 7 shows two example IDIs participants created that augmented with additional interactivity beyond the real world. (a) `Interactive Statue': the reconstructed violinist statue augmented with its beyond real-world functions of playing music by adding a ‘Play’ button and embedded content. (b) `Interactive Photo Album': the reconstructed souvenir augmented with its real-world functions of displaying photos by adding a button widget and a screen widget.}
\label{fig:creative}
\end{figure}

\subsubsection{Physical Interactivity Created by InteRecon for Realistic Experiences}
\label{fi:Realistic}
All participants agreed that they were able to create the IDI from physical items and reconstruct interactivity for them within InteRecon in the AR environment, highlighting its \engquote{memorability} (P1), and \engquote{digital longevity} (P4) of IDI. 
\revision{Additionally, with the realistic interactions of IDI created by InteRecon, participants felt a strong sense of ownership and envisioned IDIs as 3D interactive digital assets, \engquote{more memorable than photos, vlogs, or static 3D scanning items} (N=13).}
\revision{Participants also mentioned that IDI represents an `exciting advancement' compared to static 3D scanned digital objects (N=15). This is because the interactivity of personal items is entirely defined, designed, and reconstructed by the participants themselves. It more comprehensively reflects the meaning of personal items by reconstructing how users engage with their personal items, which are closely tied to personal memories. As a result, \engquote{a deeper emotional connection to IDI was built} (N=10).}


\textbf{Interfaces and content} reconstructed in IDI for electronic devices is vivid and relatively authentic, facilitating access to the original versions of files and allowing for the original interactive input methods of the devices.
It's akin to an AR emulator of vintage electronic devices, offering a more realistic experience than 2D emulators which lack the ability to replicate physical interactions. 
\engquote{For instance, a 2D emulator merely converts a physical button press into a screen click, diminishing its authenticity. } (P6), as he reconstructed his Game Boy in AR in Fig. \ref{fig:results}(b-2). 
After reconstructing more electronic devices (e.g., camera, music player, and DJ table) shown in Fig. \ref{fig:results}(b) and Fig. \ref{fig:teaser}(b), participants suggested that future collections of various iterations of electronic devices (like the diverse models of music players and cameras) might be supplanted by their AR counterparts, alleviating the demands on physical storage space and financial expenses. 

\textbf{Geometric properties} reconstructed in IDI for physical artifacts enhanced the enjoyment of interacting with digital replicas of memory artifacts.
Many participants (N=15) emphasized that the reconstruction of the mechanical joints truly constituted the impressive physical features of their toys, \engquote{turning them into more functional and delightful digital keepsakes.} (P9).
However, more physical properties (e.g. texture, softness, opacity, etc. ) were mentioned by our participants to improve the realism of IDI. 
For example, as Fig. \ref{fig:results}(a-2) illustrated, P10 said \engquote{I find my toy soft and furry texture very comforting, yet its IDI version appears stiff, reducing our emotional connection with it.}

\subsubsection{Immersive Interactions Empowered IDI creation}
\label{fi:immersive}
InteRecon enabled various virtual and physical interactions, facilitating users in reconstructing the physical interactivity of memory artifacts. 
Within this spectrum, two types of prominent interactions were frequently highlighted by users as significantly impacting the reconstruction experience.

\textbf{Hand-object interactions} in AR were considered the principal resource of realism. 
Many participants (N=12) expressed their appreciation for being able to use their hands to interact with virtual objects in AR as if they were in the physical world. 
As P7 mentioned in Fig. \ref{fig:teaser}(a), \engquote{When I slightly touched the digital toy Stitch, its head moved just like the real toy Stitch.} 
Bare-hand manipulation and its corresponding realistic transformation effects in IDI constitute the most important and preferable characteristic of InteRecon.
Seven participants with VR/AR experience noted that interactions simulating bare-hand physics diverge from their usual experiences in immersive environments, which typically \engquote{involve selecting or manipulating targets with a controller} (P12).
Five participants suggested further enhancements in modeling different hand parts (e.g., finger joints, palm or back of the hand, etc.) to facilitate more detailed collisions and touch interactions with IDI.
\engquote{Since interactions between various hand surfaces and objects often result in different components' movements of the object in the real world} (P14), these advancements could also increase the interaction's realism.

\textbf{Model segmentation} is considered to have raised concerns due to being overly realistic.
Two different operations were provided for users to segment 3D models, including direct segmentation in an AR environment and segmentation using 2D software.
Within the former operation, users could witness the model being divided into components. 
Many participants expressed discomfort with this effect, describing it as cruel: \engquote{Perhaps I don't want my toy to be divided into parts; it looks as if it is broken. Maybe segmenting in a 2D interface could mitigate the realism, making it more acceptable to me.} (P15).
Especially \engquote{models of animals or human figures}, can feel overly realistic and thus \engquote{cruel} (N=13)—for example, separating the legs of a toy dog or the arms of a LEGO figure, as their IDI illustrated in Fig. \ref{fig:geo} and Fig. \ref{fig:results}(a-4). 
Although participants expressed satisfaction with the final effects of IDI to enable interactivity, they did not hope the creating process was also too realistic.  
In contrast, segmentation within a 2D environment does not evoke these sentiments. 
In 2D, the process is perceived as \engquote{a routine operation without a sense of realism} (P9), thereby avoiding feelings of cruelty.

\subsubsection{Potentials of IDI to Enrich Memory Archives}
\label{fi:potentials}
Participants praised the concept of IDI in terms of enriching personal memory archives and proposed new envisioned features and scenarios to enrich the design space of InteRecon.
\revision{Participants compared the difference between simply scanning objects and objects with authoring their interactivity (N=10). They mentioned that it is similar to the difference between photos and videos; dynamic, interactive 3D models can be more expressive in life-logging scenarios. \engquote{They can be dynamic and offer greater potential for secondary creation for people} (P8), and also mentioned that \engquote{the interaction between objects, items, and people together contributes to the completeness of memories.} (P11).
}

\textbf{Creative interactivity beyond the real world} could be created by InteRecon. 
These interactions might not happen in the real world, but InteRecon can realize them in AR.
For example, for personal items, many participants wanted to create interactions about contextual elements associated with the memory (e.g., music, animations, photos, etc.) to enrich the memory of the item due to the convenience of interactivity creation by InteRecon, as shown in Fig. \ref{fig:creative}.
P4 created an IDI that could play music by pressing a `Play' button for a statue of a violinist in Mexico, as illustrated in Fig. \ref{fig:creative}(a). 
P4 added \engquote{When I bought this statue in Mexico, the environmental music was always `Remember me.' So, it is fantastic for me to reconstruct the music in my memory to my virtual statue by attaching a button.}
Also, P16 developed a feature that attaches a button widget and a screen widget to a souvenir purchased during a trip, which is shown in Fig. \ref{fig:creative}(b). This setup is designed to display photos from the trip, allowing for a natural recollection of travel memories each time this IDI is accessed, combining photos and the model.
By using InteRecon, participants could conveniently create the IDI's interactivity beyond its physical counterpart and delineate a more interactive virtual reconstruction that enriches personal memory archives.

\textbf{In-situ and life-logging scenarios} were also proposed by our participants, including using InteRecon to enable in-situ 3D interactivity reconstruction and utilizing IDI to be a social media platform to empower life-logging.
Non-personal items or items not at hand were proposed by participants to conduct in-situ interactivity reconstruction within InteRecon, such as \engquote{museum exhibits} (P8), \engquote{interactive art installations encountered during travel} and \engquote{my toy which is in my parent's house} (P4).
These items, though integral to their memories, are not physically transportable.
By using InteRecon remotely, anytime and anywhere, people could get access to their memorable items without spatial and temporal limitations, thus extending the digital longevity of the items.
Additionally, InteRecon was recognized as a life-logging 3D content generator. 
As P5 said, \engquote{3D digital replication with interactivity offers a more vivid representation than static 2D photos. Thus I can capture memorable moments with my pets by incorporating interactive elements into our digital counterparts!}.
As life-logging 3D content thrives, InteRecon has the potential to evolve into a platform for social media sharing, as P5 said \engquote{like a 3D version of Instagram}, empowering the pervasive access of 3D content.

\subsubsection{\revision{Broader Conceptual Enrichment and Applications of IDI}}
\label{broader}
\revision{Our participants also shared ideas for the future applications of IDI, enriching its conceptual framework across various professional fields.
First, IDIs can serve as instant sharing objects in both museum and educational settings.
In museums, IDI can function as an AR digital object that integrates the historical use of ancient artifacts or brings ancient sculptures to life with dynamic animations like \engquote{making Terracotta Army of China come alive} (P3). This interactivity can make exhibits more engaging and impressive for visitors.}
\revision{In educational settings, IDI acts as a medium for instant sharing, allowing teachers to demonstrate key mathematical and physical concepts like \engquote{celestial movements} or visualize literary works by incorporating interactive scenes like \engquote{realizing Harry Potter's magic} (P2).}
\revision{In the medical field, doctors can develop IDIs that provide detailed operating procedures of \engquote{cosmetic surgery, organ transplant surgery, or traditional Chinese acupuncture} (P9). 
This comprehensive presentation allows patients to gain a clearer understanding of upcoming surgeries, while also have the potential to enable apprentice doctors to enhance their knowledge for performing these procedures. 
In the field of fashion design, our participants envisioned that IDI could offer a cost-effective way to model and experiment with materials and environments. This method allows designers to preview how garments under various conditions that are difficult to replicate in the physical world. As P7 noted, \engquote{I can reconstruct my designed coat as an IDI and even see how it performs when worn in the low gravity environment of the moon!}. Overall, the rich interactive reconstruction and customization capabilities of IDI make it highly applicable across various industries.}

    \section{Discussion}
\subsection{Value and Challenges of Reconstructing Interactivity in Virtual Environments}
Physical items and their interactivity play an important role in human memory, serving as memory triggers and markers \cite{10.1145/1806923.1806924,petrelli2010family,10.1145/3024969.3024996}, which cannot be effectively captured by photos or videos. 
Our user study reveals that IDI can fill this gap, recording physical interactivity for personal items as a novel representation to enrich personal memory.
While IDI can't fully be an exact equivalent of physical items due to the lack of tactile aspects, it still offers a high level of realism by preserving interactivity. 
With this realism enabled, IDI is no longer a "cold and lifeless" virtual artifact.
Instead, it becomes a dynamic, cyber personal asset owned by users, fostering a stronger emotional connection between users and virtual objects compared to traditional recording formats like photos or videos.

Our findings also indicate that the reconstruction quality of IDI in terms of realism, influences participants' perceptions of ownership over IDI.
This aligns with the existing research, which has established a connection between the quality of 3D object reconstruction and human perceptions \cite{scarfe2015using, crete2012reconstructing, 10.1145/1518701.1518966}.
For example, participants tended to form a strong perception of ownership and connection to IDIs (Sec. \ref{fi:Realistic}) when they had been reconstructed with high fidelity, and viewed the IDI as an extension of their personal belongings.
This perception can be enhanced by incorporating more fine-grained reconstruction functions into InteRecon (Sec. \ref{fi:Realistic}).


Interactivity in the real world is multidimensional, encompassing interactions between people, objects, and environments, forming a non-parametric and complex structure that is difficult to describe with existing methods \cite{gibson1977theory,10.1145/3446370}. While IDI took the first step to identifying some interactivity features of memorable personal items, defining, creating, and editing interactivity in broader virtual environments like the metaverse remains an open question. This challenge not only requires technical support but also involves context, design, and psychological considerations. Interactivity in virtual worlds must dynamically adapt to user behavior and the environment, potentially combining user input and machine learning to create a more flexible, natural, and user-aligned interactive experience.

\subsection{IDI Ecosystem and Application Scenarios}
\subsubsection{IDI as Storage and Dissemination Tool for Personal Memory Archives}
In our user study, participants showed great interest in IDI for its potential and benefits in reconstructing personalized content. As mentioned by some participants, the use of IDI can be extended to represent, store, and disseminate personal memories and experiences associated with personal items.
Key elements like people, scenes, objects, and their interactions are valuable information to be recorded for high-fidelity memory preservation. 
Although previous work has explored using MR to rebuild these elements, such as reconstructing human figures \cite{10.1145/3139131.3139156}, reconstructing human-object interactions in office settings \cite{10.1145/3491102.3501836}, and capturing key contexts from travel \cite{10.1145/3613904.3642320}, a unified representation form is needed to organize all these multi-source data, where IDI could be a good fit.
Based on IDI representations, research questions about how to reconstruct human figures in personal memories without triggering the uncanny valley effect, how to provide more user-friendly and editable 3D personal memory reconstruction interfaces, and how to enable multi-user collaboration and cross-user dissemination to enhance reconstruction are all valuable for further research.

\subsubsection{\revision{Richer IDI Conceptual Extensions}}
\revision{While IDI is initially designed to facilitate personal memory archiving, it can be extended to become a generalized digital interactive asset across various user groups and scenarios. 
Beyond personal items, users can reconstruct a wide range of objects by adhering to the IDI workflow.
The core of IDI is its focus on the interactivity of objects, a vital concept that significantly impacts daily life.
This interactivity can encompass the characteristics, uses, and affordance of objects, as well as richer contexts like the people, environment, and emotions associated with the object.
For example, meaningful documents or heirlooms' interactivity may lie in the stories behind them about the time, place, or people connected to them.
By transforming them into IDIs, individuals can preserve these stories anchored to the object and extend their presence in the digital realm.
Echoing past research on sharing physical environments in broader contexts \cite{10.1145/3654777.3676377,10.1145/3672539.3686740,10.1145/3654777.3676379}, beyond the aim of recording personal memories, our participants also found reconstructing interactivity meets a wider range of user needs, such as instant and interpretable sharing, effective instruction, cost-efficient simulation, or even entertainment (Sec. \ref{broader}).
}
\revision{With such wide-ranging applications, }IDIs offer opportunities to form a decentralized community where users, rather than professional modelers, can create and accumulate more personalized and interactive 3D assets in the future. 
While there are some websites (e.g., sketchfab\footnote{sketchfab: https://sketchfab.com/feed}) for experts to upload or share their 3D models, there is no existing prevalent online 3D asset platform tailored for non-expert users. 
\revision{Therefore, future work can focus on establishing such a community and on designing and implementing systems for user-created virtual interactive 3D asset sharing and communication. }



\revision{\subsection{Humane Considerations with Sentimental Value for Digital Item Creation}}
We believe that developing a 3D reconstruction tool for end-users requires providing both intuitive and humane operations. 
Participants find the functions provided by InteRecon to be easy and effortless to use, and the framework and workflow are complete for reconstructing physical interactivity.
Specifically for intuitive operations, InteRecon realized by offering previews and templates for users to map, while giving them editing freedom.
We envision that InteRecon should enable more intuitive operations in the future. 
A good example is demonstration-based programming \cite{10.1145/3472749.3474769,lu2013gesture}, enabling users to define an interaction by just demonstrating this interaction.
Intuitive interactions are to lower the barrier to using InteRecon, making 3D interactive reconstruction as simple and easy as taking a photo. 

\revision{
However, intuitive operations in 3D environment may cause sentimental hurt due to its realistic effects. 
Since the reconstructed object holds significant personal memories and value for users, they may experience emotional disturbances, such as feelings of discomfort or cruelty, while reconstructing it in a 3D environment (Sec. \ref{fi:immersive}). These feelings are particularly common during operations like segmentation, splicing, deformation, and color alteration. Additionally, hyper-realistic or imperfectly replicated objects that appear temporarily during the process can evoke eerie emotions.
Therefore, we believe that humane designs regarding sentimental value should be integrated throughout the entire interactive process of creating personal digital items. 

An alternative approach of using the 2D interface for editing is mentioned in our user study, which aligns with previous research \cite{seymour2021have,mcmahan2016interaction} and may help reduce emotional disturbances. From this observation, we suggest that the pre-editing process for a 3D model, which might involve disruptive changes for 3d models, can be done in a less immersive setting: In a 2D interface with grid lines and toolboxes, the immersion is reduced, allowing users to focus more on editing without the realistic sensation of harming a cherished item.
However, manipulating a 3D model in a 2D interface may not be as intuitive as in a 3D environment. Therefore, this still needs empirical studies to explore how to seamlessly integrate 2D intermediary steps in the future. The goal is to provide users with a more intuitive and also emotionally acceptable experience when their cherished items are converted into a virtual format.
}

\subsection{Incorporating Broader Technical Development}
InteRecon successfully captures basic physical motions and mechanical structures associated with a variety of physical objects. 
Yet, it is imperative to broaden this scope to encompass more complex and varied physical effects. 
\revision{For the reconstruction fidelity, material generation models (e.g., ControlMat \cite{10.1145/3688830}, MatFuse \cite{Vecchio_2024}, TileGen \cite{zhou2022tilegentileablecontrollablematerial}) can be integrated to process complex material reconstruction (e.g., metal, leather, clay). }
As for physical transforms, despite the capability of defining a close set of common joints, future work could generalize to reconstruct physical deformations and motions at different dimensions (e.g., more complex mechanical structures such as pulley blocks and difference types of force such as torque and friction) and scales (e.g., segments with different sizes and resolutions) by incorporating advanced machine learning models \cite{ ao2023gesturediffuclip, lesser2022loki}. 
\revision{For physical-based human-object interaction authoring, Multi-modal Large Language Model (MLLM) combined with physics-based simulation can be integrated to automatically predict 3D objects' interactive dynamics under forces (e.g., SimAnything \cite{zhao2024automated3dphysicalsimulation}, DreamPhysics \cite{huang2024dreamphysicslearningphysicalproperties}, DreamGaussian4D \cite{ren2024dreamgaussian4dgenerative4dgaussian}). Thus, future work in this area can explore interaction capabilities such as dynamic controlling of flexibility, exploring thermal properties of 3D models, dynamic linkages with existing objects in an environment~\cite{stemasov2020mix,10.1145/3654777.3676379}, etc.}

InteRecon's functions of reconstructing electronic device interfaces has been initially set to basic interactions within widgets such as buttons, knobs, screens, and more mechanical triggers. 
To extend the utility and applicability of InteRecon, we propose enhancing the system to support a wider array of electrical functions and components. 
This expansion would involve integrating more complex and diverse electronic mechanisms, allowing users to simulate and interact with a broader range of device interfaces. 
For instance, incorporating touch-sensitive screens, sensor-based interactions, and advanced control systems \cite{tatzgern2022airres, liao2022realitytalk, kim2022spinocchietto} could significantly enrich user experiences. 
By enabling more sophisticated electrical interactions, users can not only recreate but also prototype new device functionalities within the InteRecon environment. 
This would not only enhance the simulation fidelity but also provide a robust platform for education, design, and prototyping activities.

\subsection{Limitations}

Our work took a further step in investigating interactivity-aware reconstruction of personal items in the MR environment, but it still has limitations that motivate potential future research directions.

Firstly, we believe integrating a more powerful physical simulation engine would improve the realism of InteRecon's hand-to-object interactions, including fine-grained simulation for furry textures or soft bodies and detailed modeling of human hand joints. 
Thus, the collision, touching, or other random interactions between hands and virtual objects will be more realistic. 
Furthermore, it is noteworthy that participants encountered initial challenges when learning to navigate the AR interface in Hololens2 by using hand gestures for selecting models or buttons, partly due to lower pinching gesture accuracy. Therefore, a more robust hand-tracking algorithm should be implemented to improve InteRecon in the future.

The creation of virtual assets (e.g., embedded content, widgets) also potentially be augmented by AI-based generation technology by text prompt \cite{song2023consistency, zhang2023amphion, liu2024sora} or image prompt \cite{podell2023sdxl, zhu2023minigpt, wang2023dreamvideo}. For example, users can leverage text or pictures to generate the corresponding 3D models if they do not have the physical item at hand. 
However, the trade-off between AI generation and the freedom of user customization needs to be further explored and validated through experiments, as personal items can contain many unique and individualized features.




    \section{Conclusion}
In this work, we presented the concept of Interactive Digital Item (IDI), emphasizing the reconstruction of memorable personal items while preserving their interactivity, which represented the design goals of reconstructing geometry, interface, and embedded content.
IDI is built on the knowledge of a formative study we conducted to understand the participants' expectations on the interactivity features of memorable personal items that wanted to be reconstructed for digital replications.
We developed an AR prototype, InteRecon, which allows users to create and present IDI within MR environments from physical items, including four main categories of functions: 1) Reconstructing 3D appearance, 2) Adding physical transforms, 3) Reconstructing interface, 4) Adding embedded content.
We conducted a two-session user study to assess the feasibility of using InteRecon to create IDI and collecting users' feedback on the challenges, future opportunities,
and applications of IDI. 
We collected quantitative and qualitative feedback and the results revealed that InteRecon is effective, expressive, and enjoyable for creating IDIs. 
Furthermore, the qualitative feedback illustrated that 1) IDIs brought realistic experiences with physical interactivity; 2) The immersive interactions enabled by InteRecon empowered the IDI creation; and 3) IDIs have the potential to enrich personal memory archives.

\begin{acks}
This work is partially supported by the Guangzhou-HKUST(GZ) Joint Funding Project (No. 2024A03J0617), Guangzhou Higher Education Teaching Quality and Teaching Reform Project (No. 2024YBJG070),  Education Bureau of Guangzhou Municipality Funding Project  (No. 2024312152), Guangdong Provincial Key Lab of Integrated Communication, Sensing and Computation for Ubiquitous Internet of Things (No. 2023B1212010007), the Project of DEGP (No.2023KCXTD042), and the Guangzhou Science and Technology Program City-University Joint Funding Project (No. 2023A03J0001).
\end{acks}

\bibliographystyle{ACM-Reference-Format}
\bibliography{main}


\begin{thebibliography}{107}


\ifx \showCODEN    \undefined \def \showCODEN     #1{\unskip}     \fi
\ifx \showDOI      \undefined \def \showDOI       #1{#1}\fi
\ifx \showISBNx    \undefined \def \showISBNx     #1{\unskip}     \fi
\ifx \showISBNxiii \undefined \def \showISBNxiii  #1{\unskip}     \fi
\ifx \showISSN     \undefined \def \showISSN      #1{\unskip}     \fi
\ifx \showLCCN     \undefined \def \showLCCN      #1{\unskip}     \fi
\ifx \shownote     \undefined \def \shownote      #1{#1}          \fi
\ifx \showarticletitle \undefined \def \showarticletitle #1{#1}   \fi
\ifx \showURL      \undefined \def \showURL       {\relax}        \fi
\providecommand\bibfield[2]{#2}
\providecommand\bibinfo[2]{#2}
\providecommand\natexlab[1]{#1}
\providecommand\showeprint[2][]{arXiv:#2}

\bibitem[Addis et~al\mbox{.}(2010)]%
        {addis2010100}
\bibfield{author}{\bibinfo{person}{Matthew Addis}, \bibinfo{person}{Walter Allasia}, \bibinfo{person}{Werner Bailer}, \bibinfo{person}{Laurent Boch}, \bibinfo{person}{Francesco Gallo}, {and} \bibinfo{person}{Richard Wright}.} \bibinfo{year}{2010}\natexlab{}.
\newblock \showarticletitle{100 million hours of audiovisual content: digital preservation and access in the PrestoPRIME project}. In \bibinfo{booktitle}{\emph{Proceedings of the 1st International Digital Preservation Interoperability Framework Symposium}}. \bibinfo{pages}{1--8}.
\newblock


\bibitem[Ao et~al\mbox{.}(2023)]%
        {ao2023gesturediffuclip}
\bibfield{author}{\bibinfo{person}{Tenglong Ao}, \bibinfo{person}{Zeyi Zhang}, {and} \bibinfo{person}{Libin Liu}.} \bibinfo{year}{2023}\natexlab{}.
\newblock \showarticletitle{Gesturediffuclip: Gesture diffusion model with clip latents}.
\newblock \bibinfo{journal}{\emph{ACM Transactions on Graphics (TOG)}} \bibinfo{volume}{42}, \bibinfo{number}{4} (\bibinfo{year}{2023}), \bibinfo{pages}{1--18}.
\newblock


\bibitem[Banks and Sellen(2009)]%
        {10.1145/1517664.1517678}
\bibfield{author}{\bibinfo{person}{Richard Banks} {and} \bibinfo{person}{Abigail Sellen}.} \bibinfo{year}{2009}\natexlab{}.
\newblock \showarticletitle{Shoebox: mixing storage and display of digital images in the home}. In \bibinfo{booktitle}{\emph{Proceedings of the 3rd International Conference on Tangible and Embedded Interaction}} (Cambridge, United Kingdom) \emph{(\bibinfo{series}{TEI '09})}. \bibinfo{publisher}{Association for Computing Machinery}, \bibinfo{address}{New York, NY, USA}, \bibinfo{pages}{35–40}.
\newblock
\showISBNx{9781605584935}
\urldef\tempurl%
\url{https://doi.org/10.1145/1517664.1517678}
\showDOI{\tempurl}


\bibitem[Beigl et~al\mbox{.}(2001)]%
        {beigl2001mediacups}
\bibfield{author}{\bibinfo{person}{Michael Beigl}, \bibinfo{person}{Hans-W Gellersen}, {and} \bibinfo{person}{Albrecht Schmidt}.} \bibinfo{year}{2001}\natexlab{}.
\newblock \showarticletitle{Mediacups: experience with design and use of computer-augmented everyday artefacts}.
\newblock \bibinfo{journal}{\emph{Computer Networks}} \bibinfo{volume}{35}, \bibinfo{number}{4} (\bibinfo{year}{2001}), \bibinfo{pages}{401--409}.
\newblock


\bibitem[Blake(1985)]%
        {blake1985design}
\bibfield{author}{\bibinfo{person}{Alexander Blake}.} \bibinfo{year}{1985}\natexlab{}.
\newblock \bibinfo{booktitle}{\emph{Design of mechanical joints}}. Vol.~\bibinfo{volume}{42}.
\newblock \bibinfo{publisher}{CRC press}.
\newblock


\bibitem[Bowen and Petrelli(2011)]%
        {bowen2011remembering}
\bibfield{author}{\bibinfo{person}{Simon Bowen} {and} \bibinfo{person}{Daniela Petrelli}.} \bibinfo{year}{2011}\natexlab{}.
\newblock \showarticletitle{Remembering today tomorrow: Exploring the human-centred design of digital mementos}.
\newblock \bibinfo{journal}{\emph{International Journal of Human-Computer Studies}} \bibinfo{volume}{69}, \bibinfo{number}{5} (\bibinfo{year}{2011}), \bibinfo{pages}{324--337}.
\newblock


\bibitem[Bradley(2014)]%
        {bradley2014emotional}
\bibfield{author}{\bibinfo{person}{Margaret~M Bradley}.} \bibinfo{year}{2014}\natexlab{}.
\newblock \showarticletitle{Emotional memory: A dimensional analysis}.
\newblock In \bibinfo{booktitle}{\emph{Emotions}}. \bibinfo{publisher}{Psychology Press}, \bibinfo{pages}{97--134}.
\newblock


\bibitem[Braun and Clarke(2006)]%
        {braun2006using}
\bibfield{author}{\bibinfo{person}{Virginia Braun} {and} \bibinfo{person}{Victoria Clarke}.} \bibinfo{year}{2006}\natexlab{}.
\newblock \showarticletitle{Using thematic analysis in psychology}.
\newblock \bibinfo{journal}{\emph{Qualitative research in psychology}} \bibinfo{volume}{3}, \bibinfo{number}{2} (\bibinfo{year}{2006}), \bibinfo{pages}{77--101}.
\newblock


\bibitem[Broekhuijsen et~al\mbox{.}(2017)]%
        {10.1145/3024969.3024996}
\bibfield{author}{\bibinfo{person}{Mendel Broekhuijsen}, \bibinfo{person}{Elise van~den Hoven}, {and} \bibinfo{person}{Panos Markopoulos}.} \bibinfo{year}{2017}\natexlab{}.
\newblock \showarticletitle{Design Directions for Media-Supported Collocated Remembering Practices}. In \bibinfo{booktitle}{\emph{Proceedings of the Eleventh International Conference on Tangible, Embedded, and Embodied Interaction}} (Yokohama, Japan) \emph{(\bibinfo{series}{TEI '17})}. \bibinfo{publisher}{Association for Computing Machinery}, \bibinfo{address}{New York, NY, USA}, \bibinfo{pages}{21–30}.
\newblock
\showISBNx{9781450346764}
\urldef\tempurl%
\url{https://doi.org/10.1145/3024969.3024996}
\showDOI{\tempurl}


\bibitem[Cai et~al\mbox{.}(2024)]%
        {10.1145/3613904.3642320}
\bibfield{author}{\bibinfo{person}{Runze Cai}, \bibinfo{person}{Nuwan Janaka}, \bibinfo{person}{Yang Chen}, \bibinfo{person}{Lucia Wang}, \bibinfo{person}{Shengdong Zhao}, {and} \bibinfo{person}{Can Liu}.} \bibinfo{year}{2024}\natexlab{}.
\newblock \showarticletitle{PANDALens: Towards AI-Assisted In-Context Writing on OHMD During Travels}. In \bibinfo{booktitle}{\emph{Proceedings of the CHI Conference on Human Factors in Computing Systems}} (Honolulu, HI, USA) \emph{(\bibinfo{series}{CHI '24})}. \bibinfo{publisher}{Association for Computing Machinery}, \bibinfo{address}{New York, NY, USA}, Article \bibinfo{articleno}{1053}, \bibinfo{numpages}{24}~pages.
\newblock
\showISBNx{9798400703300}
\urldef\tempurl%
\url{https://doi.org/10.1145/3613904.3642320}
\showDOI{\tempurl}


\bibitem[Chen et~al\mbox{.}(2019)]%
        {10.1145/3322276.3322301}
\bibfield{author}{\bibinfo{person}{Amy Yo~Sue Chen}, \bibinfo{person}{William Odom}, \bibinfo{person}{Ce Zhong}, \bibinfo{person}{Henry Lin}, {and} \bibinfo{person}{Tal Amram}.} \bibinfo{year}{2019}\natexlab{}.
\newblock \showarticletitle{Chronoscope: Designing Temporally Diverse Interactions with Personal Digital Photo Collections}. In \bibinfo{booktitle}{\emph{Proceedings of the 2019 on Designing Interactive Systems Conference}} (San Diego, CA, USA) \emph{(\bibinfo{series}{DIS '19})}. \bibinfo{publisher}{Association for Computing Machinery}, \bibinfo{address}{New York, NY, USA}, \bibinfo{pages}{799–812}.
\newblock
\showISBNx{9781450358507}
\urldef\tempurl%
\url{https://doi.org/10.1145/3322276.3322301}
\showDOI{\tempurl}


\bibitem[Chen et~al\mbox{.}(2022)]%
        {9879997}
\bibfield{author}{\bibinfo{person}{Hsiao-yu Chen}, \bibinfo{person}{Edith Tretschk}, \bibinfo{person}{Tuur Stuyck}, \bibinfo{person}{Petr Kadlecek}, \bibinfo{person}{Ladislav Kavan}, \bibinfo{person}{Etienne Vouga}, {and} \bibinfo{person}{Christoph Lassner}.} \bibinfo{year}{2022}\natexlab{}.
\newblock \showarticletitle{Virtual Elastic Objects}. In \bibinfo{booktitle}{\emph{2022 IEEE/CVF Conference on Computer Vision and Pattern Recognition (CVPR)}}. \bibinfo{pages}{15806--15816}.
\newblock
\urldef\tempurl%
\url{https://doi.org/10.1109/CVPR52688.2022.01537}
\showDOI{\tempurl}


\bibitem[Crabtree et~al\mbox{.}(2004)]%
        {10.1145/1031607.1031673}
\bibfield{author}{\bibinfo{person}{Andy Crabtree}, \bibinfo{person}{Tom Rodden}, {and} \bibinfo{person}{John Mariani}.} \bibinfo{year}{2004}\natexlab{}.
\newblock \showarticletitle{Collaborating around collections: informing the continued development of photoware}. In \bibinfo{booktitle}{\emph{Proceedings of the 2004 ACM Conference on Computer Supported Cooperative Work}} (Chicago, Illinois, USA) \emph{(\bibinfo{series}{CSCW '04})}. \bibinfo{publisher}{Association for Computing Machinery}, \bibinfo{address}{New York, NY, USA}, \bibinfo{pages}{396–405}.
\newblock
\showISBNx{1581138105}
\urldef\tempurl%
\url{https://doi.org/10.1145/1031607.1031673}
\showDOI{\tempurl}


\bibitem[Crete-Nishihata et~al\mbox{.}(2012)]%
        {crete2012reconstructing}
\bibfield{author}{\bibinfo{person}{Masashi Crete-Nishihata}, \bibinfo{person}{Ronald~M Baecker}, \bibinfo{person}{Michael Massimi}, \bibinfo{person}{Deborah Ptak}, \bibinfo{person}{Rachelle Campigotto}, \bibinfo{person}{Liam~D Kaufman}, \bibinfo{person}{Adam~M Brickman}, \bibinfo{person}{Gary~R Turner}, \bibinfo{person}{Joshua~R Steinerman}, {and} \bibinfo{person}{Sandra~E Black}.} \bibinfo{year}{2012}\natexlab{}.
\newblock \showarticletitle{Reconstructing the past: personal memory technologies are not just personal and not just for memory}.
\newblock \bibinfo{journal}{\emph{Human--Computer Interaction}} \bibinfo{volume}{27}, \bibinfo{number}{1-2} (\bibinfo{year}{2012}), \bibinfo{pages}{92--123}.
\newblock


\bibitem[Cushing(2011)]%
        {cushing2011self}
\bibfield{author}{\bibinfo{person}{Amber~L Cushing}.} \bibinfo{year}{2011}\natexlab{}.
\newblock \showarticletitle{Self extension and the desire to preserve digital possessions}.
\newblock \bibinfo{journal}{\emph{Proceedings of the American Society for Information Science and Technology}} \bibinfo{volume}{48}, \bibinfo{number}{1} (\bibinfo{year}{2011}), \bibinfo{pages}{1--3}.
\newblock


\bibitem[Dobbins et~al\mbox{.}(2014)]%
        {dobbins2014creating}
\bibfield{author}{\bibinfo{person}{Chelsea Dobbins}, \bibinfo{person}{Madjid Merabti}, \bibinfo{person}{Paul Fergus}, {and} \bibinfo{person}{David Llewellyn-Jones}.} \bibinfo{year}{2014}\natexlab{}.
\newblock \showarticletitle{Creating human digital memories with the aid of pervasive mobile devices}.
\newblock \bibinfo{journal}{\emph{Pervasive and Mobile Computing}}  \bibinfo{volume}{12} (\bibinfo{year}{2014}), \bibinfo{pages}{160--178}.
\newblock


\bibitem[Dogan et~al\mbox{.}(2024)]%
        {10.1145/3654777.3676379}
\bibfield{author}{\bibinfo{person}{Mustafa~Doga Dogan}, \bibinfo{person}{Eric~J Gonzalez}, \bibinfo{person}{Karan Ahuja}, \bibinfo{person}{Ruofei Du}, \bibinfo{person}{Andrea Cola\c{c}o}, \bibinfo{person}{Johnny Lee}, \bibinfo{person}{Mar Gonzalez-Franco}, {and} \bibinfo{person}{David Kim}.} \bibinfo{year}{2024}\natexlab{}.
\newblock \showarticletitle{Augmented Object Intelligence with XR-Objects}. In \bibinfo{booktitle}{\emph{Proceedings of the 37th Annual ACM Symposium on User Interface Software and Technology}} (Pittsburgh, PA, USA) \emph{(\bibinfo{series}{UIST '24})}. \bibinfo{publisher}{Association for Computing Machinery}, \bibinfo{address}{New York, NY, USA}, Article \bibinfo{articleno}{19}, \bibinfo{numpages}{15}~pages.
\newblock
\showISBNx{9798400706288}
\urldef\tempurl%
\url{https://doi.org/10.1145/3654777.3676379}
\showDOI{\tempurl}


\bibitem[Duranti and Shaffer(2012)]%
        {duranti2012memory}
\bibfield{author}{\bibinfo{person}{Luciana Duranti} {and} \bibinfo{person}{Elizabeth Shaffer}.} \bibinfo{year}{2012}\natexlab{}.
\newblock \showarticletitle{The memory of the world in the digital age: digitization and preservation}. In \bibinfo{booktitle}{\emph{An International Conference on Permanent Access to Digital Documentary Heritage, UNESCO Conference Proceedings. Vancouver}}.
\newblock


\bibitem[Faruqi et~al\mbox{.}(2023)]%
        {10.1145/3586183.3606723}
\bibfield{author}{\bibinfo{person}{Faraz Faruqi}, \bibinfo{person}{Ahmed Katary}, \bibinfo{person}{Tarik Hasic}, \bibinfo{person}{Amira Abdel-Rahman}, \bibinfo{person}{Nayeemur Rahman}, \bibinfo{person}{Leandra Tejedor}, \bibinfo{person}{Mackenzie Leake}, \bibinfo{person}{Megan Hofmann}, {and} \bibinfo{person}{Stefanie Mueller}.} \bibinfo{year}{2023}\natexlab{}.
\newblock \showarticletitle{Style2Fab: Functionality-Aware Segmentation for Fabricating Personalized 3D Models with Generative AI}. In \bibinfo{booktitle}{\emph{Proceedings of the 36th Annual ACM Symposium on User Interface Software and Technology}} (San Francisco, CA, USA) \emph{(\bibinfo{series}{UIST '23})}. \bibinfo{publisher}{Association for Computing Machinery}, \bibinfo{address}{New York, NY, USA}, Article \bibinfo{articleno}{22}, \bibinfo{numpages}{13}~pages.
\newblock
\showISBNx{9798400701320}
\urldef\tempurl%
\url{https://doi.org/10.1145/3586183.3606723}
\showDOI{\tempurl}


\bibitem[Feinberg(2013)]%
        {10.1145/2470654.2466453}
\bibfield{author}{\bibinfo{person}{Melanie Feinberg}.} \bibinfo{year}{2013}\natexlab{}.
\newblock \showarticletitle{Beyond Digital and Physical Objects: The Intellectual Work as a Concept of Interest for HCI}. In \bibinfo{booktitle}{\emph{Proceedings of the SIGCHI Conference on Human Factors in Computing Systems}} (Paris, France) \emph{(\bibinfo{series}{CHI '13})}. \bibinfo{publisher}{Association for Computing Machinery}, \bibinfo{address}{New York, NY, USA}, \bibinfo{pages}{3317–3326}.
\newblock
\showISBNx{9781450318990}
\urldef\tempurl%
\url{https://doi.org/10.1145/2470654.2466453}
\showDOI{\tempurl}


\bibitem[Fender and Holz(2022)]%
        {10.1145/3491102.3501836}
\bibfield{author}{\bibinfo{person}{Andreas~Rene Fender} {and} \bibinfo{person}{Christian Holz}.} \bibinfo{year}{2022}\natexlab{}.
\newblock \showarticletitle{Causality-preserving Asynchronous Reality}. In \bibinfo{booktitle}{\emph{Proceedings of the 2022 CHI Conference on Human Factors in Computing Systems}} (<conf-loc>, <city>New Orleans</city>, <state>LA</state>, <country>USA</country>, </conf-loc>) \emph{(\bibinfo{series}{CHI '22})}. \bibinfo{publisher}{Association for Computing Machinery}, \bibinfo{address}{New York, NY, USA}, Article \bibinfo{articleno}{634}, \bibinfo{numpages}{15}~pages.
\newblock
\showISBNx{9781450391573}
\urldef\tempurl%
\url{https://doi.org/10.1145/3491102.3501836}
\showDOI{\tempurl}


\bibitem[Garde-Hansen(2011)]%
        {garde2011digital}
\bibfield{author}{\bibinfo{person}{J Garde-Hansen}.} \bibinfo{year}{2011}\natexlab{}.
\newblock \showarticletitle{Digital Memories: The Democratisation of Archives}.
\newblock \bibinfo{journal}{\emph{Media and Memory}} (\bibinfo{year}{2011}), \bibinfo{pages}{70--88}.
\newblock


\bibitem[Gibson(1977)]%
        {gibson1977theory}
\bibfield{author}{\bibinfo{person}{James~J Gibson}.} \bibinfo{year}{1977}\natexlab{}.
\newblock \showarticletitle{The theory of affordances}.
\newblock \bibinfo{journal}{\emph{Hilldale, USA}} \bibinfo{volume}{1}, \bibinfo{number}{2} (\bibinfo{year}{1977}), \bibinfo{pages}{67--82}.
\newblock


\bibitem[Golsteijn et~al\mbox{.}(2012)]%
        {10.1145/2317956.2318054}
\bibfield{author}{\bibinfo{person}{Connie Golsteijn}, \bibinfo{person}{Elise van~den Hoven}, \bibinfo{person}{David Frohlich}, {and} \bibinfo{person}{Abigail Sellen}.} \bibinfo{year}{2012}\natexlab{}.
\newblock \showarticletitle{Towards a more cherishable digital object}. In \bibinfo{booktitle}{\emph{Proceedings of the Designing Interactive Systems Conference}} (Newcastle Upon Tyne, United Kingdom) \emph{(\bibinfo{series}{DIS '12})}. \bibinfo{publisher}{Association for Computing Machinery}, \bibinfo{address}{New York, NY, USA}, \bibinfo{pages}{655–664}.
\newblock
\showISBNx{9781450312103}
\urldef\tempurl%
\url{https://doi.org/10.1145/2317956.2318054}
\showDOI{\tempurl}


\bibitem[Habermas and Paha(2002)]%
        {habermas2002souvenirs}
\bibfield{author}{\bibinfo{person}{Tilmann Habermas} {and} \bibinfo{person}{Christine Paha}.} \bibinfo{year}{2002}\natexlab{}.
\newblock \showarticletitle{Souvenirs and other personal objects: Reminding of past events and significant others in the transition to university}.
\newblock \bibinfo{journal}{\emph{Critical advances in reminiscence work}} (\bibinfo{year}{2002}), \bibinfo{pages}{123--138}.
\newblock


\bibitem[H\"{a}kkil\"{a} et~al\mbox{.}(2019)]%
        {10.1145/3365610.3368425}
\bibfield{author}{\bibinfo{person}{Jonna H\"{a}kkil\"{a}}, \bibinfo{person}{Petri Hannula}, \bibinfo{person}{Elina Luiro}, \bibinfo{person}{Emilia Launne}, \bibinfo{person}{Sanni Mustonen}, \bibinfo{person}{Toni Westerlund}, {and} \bibinfo{person}{Ashley Colley}.} \bibinfo{year}{2019}\natexlab{}.
\newblock \showarticletitle{Visiting a virtual graveyard: designing virtual reality cultural heritage experiences}. In \bibinfo{booktitle}{\emph{Proceedings of the 18th International Conference on Mobile and Ubiquitous Multimedia}} (Pisa, Italy) \emph{(\bibinfo{series}{MUM '19})}. \bibinfo{publisher}{Association for Computing Machinery}, \bibinfo{address}{New York, NY, USA}, Article \bibinfo{articleno}{56}, \bibinfo{numpages}{4}~pages.
\newblock
\showISBNx{9781450376242}
\urldef\tempurl%
\url{https://doi.org/10.1145/3365610.3368425}
\showDOI{\tempurl}


\bibitem[Hassanin et~al\mbox{.}(2021)]%
        {10.1145/3446370}
\bibfield{author}{\bibinfo{person}{Mohammed Hassanin}, \bibinfo{person}{Salman Khan}, {and} \bibinfo{person}{Murat Tahtali}.} \bibinfo{year}{2021}\natexlab{}.
\newblock \showarticletitle{Visual Affordance and Function Understanding: A Survey}.
\newblock \bibinfo{journal}{\emph{ACM Comput. Surv.}} \bibinfo{volume}{54}, \bibinfo{number}{3}, Article \bibinfo{articleno}{47} (\bibinfo{date}{apr} \bibinfo{year}{2021}), \bibinfo{numpages}{35}~pages.
\newblock
\showISSN{0360-0300}
\urldef\tempurl%
\url{https://doi.org/10.1145/3446370}
\showDOI{\tempurl}


\bibitem[Hawkins et~al\mbox{.}(2015)]%
        {hawkins2015postulater}
\bibfield{author}{\bibinfo{person}{Daniel Hawkins}, \bibinfo{person}{Carman Neustaedter}, {and} \bibinfo{person}{Jason Procyk}.} \bibinfo{year}{2015}\natexlab{}.
\newblock \showarticletitle{Postulater: the design and evaluation of a time-delayed media sharing system}. In \bibinfo{booktitle}{\emph{Proceedings of the 41st Graphics Interface Conference}}. \bibinfo{pages}{249--256}.
\newblock


\bibitem[He et~al\mbox{.}(2023)]%
        {10.1145/3544548.3580704}
\bibfield{author}{\bibinfo{person}{Fengming He}, \bibinfo{person}{Xiyun Hu}, \bibinfo{person}{Jingyu Shi}, \bibinfo{person}{Xun Qian}, \bibinfo{person}{Tianyi Wang}, {and} \bibinfo{person}{Karthik Ramani}.} \bibinfo{year}{2023}\natexlab{}.
\newblock \showarticletitle{Ubi Edge: Authoring Edge-Based Opportunistic Tangible User Interfaces in Augmented Reality}. In \bibinfo{booktitle}{\emph{Proceedings of the 2023 CHI Conference on Human Factors in Computing Systems}} (Hamburg, Germany) \emph{(\bibinfo{series}{CHI '23})}. \bibinfo{publisher}{Association for Computing Machinery}, \bibinfo{address}{New York, NY, USA}, Article \bibinfo{articleno}{461}, \bibinfo{numpages}{14}~pages.
\newblock
\showISBNx{9781450394215}
\urldef\tempurl%
\url{https://doi.org/10.1145/3544548.3580704}
\showDOI{\tempurl}


\bibitem[Hilliges and Kirk(2009)]%
        {hilliges2009getting}
\bibfield{author}{\bibinfo{person}{Otmar Hilliges} {and} \bibinfo{person}{David~Stanley Kirk}.} \bibinfo{year}{2009}\natexlab{}.
\newblock \showarticletitle{Getting sidetracked: display design and occasioning photo-talk with the photohelix}. In \bibinfo{booktitle}{\emph{Proceedings of the SIGCHI Conference on Human Factors in Computing Systems}}. \bibinfo{pages}{1733--1736}.
\newblock


\bibitem[Hoskins(2017)]%
        {hoskins2017restless}
\bibfield{author}{\bibinfo{person}{Andrew Hoskins}.} \bibinfo{year}{2017}\natexlab{}.
\newblock \showarticletitle{The restless past: An introduction to digital memory and media}.
\newblock In \bibinfo{booktitle}{\emph{Digital memory studies}}. \bibinfo{publisher}{Routledge}, \bibinfo{pages}{1--24}.
\newblock


\bibitem[Hu et~al\mbox{.}(2023)]%
        {10.1145/3544548.3581148}
\bibfield{author}{\bibinfo{person}{Erzhen Hu}, \bibinfo{person}{Jens Emil~Sloth Gr\o{}nb\ae{}k}, \bibinfo{person}{Wen Ying}, \bibinfo{person}{Ruofei Du}, {and} \bibinfo{person}{Seongkook Heo}.} \bibinfo{year}{2023}\natexlab{}.
\newblock \showarticletitle{ThingShare: Ad-Hoc Digital Copies of Physical Objects for Sharing Things in Video Meetings}. In \bibinfo{booktitle}{\emph{Proceedings of the 2023 CHI Conference on Human Factors in Computing Systems}} (Hamburg, Germany) \emph{(\bibinfo{series}{CHI '23})}. \bibinfo{publisher}{Association for Computing Machinery}, \bibinfo{address}{New York, NY, USA}, Article \bibinfo{articleno}{365}, \bibinfo{numpages}{22}~pages.
\newblock
\showISBNx{9781450394215}
\urldef\tempurl%
\url{https://doi.org/10.1145/3544548.3581148}
\showDOI{\tempurl}


\bibitem[Hu et~al\mbox{.}(2024)]%
        {10.1145/3672539.3686740}
\bibfield{author}{\bibinfo{person}{Erzhen Hu}, \bibinfo{person}{Mingyi Li}, \bibinfo{person}{Xun Qian}, \bibinfo{person}{Alex Olwal}, \bibinfo{person}{David Kim}, \bibinfo{person}{Seongkook Heo}, {and} \bibinfo{person}{Ruofei Du}.} \bibinfo{year}{2024}\natexlab{}.
\newblock \showarticletitle{Experiencing Thing2Reality: Transforming 2D Content into Conditioned Multiviews and 3D Gaussian Objects for XR Communication}. In \bibinfo{booktitle}{\emph{Adjunct Proceedings of the 37th Annual ACM Symposium on User Interface Software and Technology}} (Pittsburgh, PA, USA) \emph{(\bibinfo{series}{UIST Adjunct '24})}. \bibinfo{publisher}{Association for Computing Machinery}, \bibinfo{address}{New York, NY, USA}, Article \bibinfo{articleno}{23}, \bibinfo{numpages}{3}~pages.
\newblock
\showISBNx{9798400707186}
\urldef\tempurl%
\url{https://doi.org/10.1145/3672539.3686740}
\showDOI{\tempurl}


\bibitem[Huang and Klippel(2020)]%
        {10.1145/3385956.3418945}
\bibfield{author}{\bibinfo{person}{Jiawei Huang} {and} \bibinfo{person}{Alexander Klippel}.} \bibinfo{year}{2020}\natexlab{}.
\newblock \showarticletitle{The Effects of Visual Realism on Spatial Memory and Exploration Patterns in Virtual Reality}. In \bibinfo{booktitle}{\emph{Proceedings of the 26th ACM Symposium on Virtual Reality Software and Technology}} (Virtual Event, Canada) \emph{(\bibinfo{series}{VRST '20})}. \bibinfo{publisher}{Association for Computing Machinery}, \bibinfo{address}{New York, NY, USA}, Article \bibinfo{articleno}{18}, \bibinfo{numpages}{11}~pages.
\newblock
\showISBNx{9781450376198}
\urldef\tempurl%
\url{https://doi.org/10.1145/3385956.3418945}
\showDOI{\tempurl}


\bibitem[Huang et~al\mbox{.}(2024b)]%
        {huang2024dreamphysicslearningphysicalproperties}
\bibfield{author}{\bibinfo{person}{Tianyu Huang}, \bibinfo{person}{Haoze Zhang}, \bibinfo{person}{Yihan Zeng}, \bibinfo{person}{Zhilu Zhang}, \bibinfo{person}{Hui Li}, \bibinfo{person}{Wangmeng Zuo}, {and} \bibinfo{person}{Rynson W.~H. Lau}.} \bibinfo{year}{2024}\natexlab{b}.
\newblock \bibinfo{title}{DreamPhysics: Learning Physical Properties of Dynamic 3D Gaussians with Video Diffusion Priors}.
\newblock
\newblock
\showeprint[arxiv]{2406.01476}~[cs.CV]
\urldef\tempurl%
\url{https://arxiv.org/abs/2406.01476}
\showURL{%
\tempurl}


\bibitem[Huang et~al\mbox{.}(2024a)]%
        {10.1145/3654777.3676377}
\bibfield{author}{\bibinfo{person}{Xincheng Huang}, \bibinfo{person}{Michael Yin}, \bibinfo{person}{Ziyi Xia}, {and} \bibinfo{person}{Robert Xiao}.} \bibinfo{year}{2024}\natexlab{a}.
\newblock \showarticletitle{VirtualNexus: Enhancing 360-Degree Video AR/VR Collaboration with Environment Cutouts and Virtual Replicas}. In \bibinfo{booktitle}{\emph{Proceedings of the 37th Annual ACM Symposium on User Interface Software and Technology}} (Pittsburgh, PA, USA) \emph{(\bibinfo{series}{UIST '24})}. \bibinfo{publisher}{Association for Computing Machinery}, \bibinfo{address}{New York, NY, USA}, Article \bibinfo{articleno}{55}, \bibinfo{numpages}{12}~pages.
\newblock
\showISBNx{9798400706288}
\urldef\tempurl%
\url{https://doi.org/10.1145/3654777.3676377}
\showDOI{\tempurl}


\bibitem[Iriye and St.~Jacques(2021)]%
        {iriye2021memories}
\bibfield{author}{\bibinfo{person}{Heather Iriye} {and} \bibinfo{person}{Peggy~L St.~Jacques}.} \bibinfo{year}{2021}\natexlab{}.
\newblock \showarticletitle{Memories for third-person experiences in immersive virtual reality}.
\newblock \bibinfo{journal}{\emph{Scientific reports}} \bibinfo{volume}{11}, \bibinfo{number}{1} (\bibinfo{year}{2021}), \bibinfo{pages}{4667}.
\newblock


\bibitem[Jansen et~al\mbox{.}(2014)]%
        {jansen2014pearl}
\bibfield{author}{\bibinfo{person}{Martijn Jansen}, \bibinfo{person}{Elise van~den Hoven}, {and} \bibinfo{person}{David Frohlich}.} \bibinfo{year}{2014}\natexlab{}.
\newblock \showarticletitle{Pearl: living media enabled by interactive photo projection}.
\newblock \bibinfo{journal}{\emph{Personal and Ubiquitous Computing}}  \bibinfo{volume}{18} (\bibinfo{year}{2014}), \bibinfo{pages}{1259--1275}.
\newblock


\bibitem[Jiang et~al\mbox{.}(2024)]%
        {10.1145/3641519.3657448}
\bibfield{author}{\bibinfo{person}{Ying Jiang}, \bibinfo{person}{Chang Yu}, \bibinfo{person}{Tianyi Xie}, \bibinfo{person}{Xuan Li}, \bibinfo{person}{Yutao Feng}, \bibinfo{person}{Huamin Wang}, \bibinfo{person}{Minchen Li}, \bibinfo{person}{Henry Lau}, \bibinfo{person}{Feng Gao}, \bibinfo{person}{Yin Yang}, {and} \bibinfo{person}{Chenfanfu Jiang}.} \bibinfo{year}{2024}\natexlab{}.
\newblock \showarticletitle{VR-GS: A Physical Dynamics-Aware Interactive Gaussian Splatting System in Virtual Reality}. In \bibinfo{booktitle}{\emph{ACM SIGGRAPH 2024 Conference Papers}} (Denver, CO, USA) \emph{(\bibinfo{series}{SIGGRAPH '24})}. \bibinfo{publisher}{Association for Computing Machinery}, \bibinfo{address}{New York, NY, USA}, Article \bibinfo{articleno}{78}, \bibinfo{numpages}{1}~pages.
\newblock
\showISBNx{9798400705250}
\urldef\tempurl%
\url{https://doi.org/10.1145/3641519.3657448}
\showDOI{\tempurl}


\bibitem[Jones and Ackerman(2018a)]%
        {jones2018co}
\bibfield{author}{\bibinfo{person}{Jasmine Jones} {and} \bibinfo{person}{Mark~S Ackerman}.} \bibinfo{year}{2018}\natexlab{a}.
\newblock \showarticletitle{Co-constructing family memory: Understanding the intergenerational practices of passing on family stories}. In \bibinfo{booktitle}{\emph{Proceedings of the 2018 chi conference on human factors in computing systems}}. \bibinfo{pages}{1--13}.
\newblock


\bibitem[Jones and Ackerman(2018b)]%
        {10.1145/3173574.3173998}
\bibfield{author}{\bibinfo{person}{Jasmine Jones} {and} \bibinfo{person}{Mark~S. Ackerman}.} \bibinfo{year}{2018}\natexlab{b}.
\newblock \showarticletitle{Co-Constructing Family Memory: Understanding the Intergenerational Practices of Passing on Family Stories}. In \bibinfo{booktitle}{\emph{Proceedings of the 2018 CHI Conference on Human Factors in Computing Systems}} (Montreal QC, Canada) \emph{(\bibinfo{series}{CHI '18})}. \bibinfo{publisher}{Association for Computing Machinery}, \bibinfo{address}{New York, NY, USA}, \bibinfo{pages}{1–13}.
\newblock
\showISBNx{9781450356206}
\urldef\tempurl%
\url{https://doi.org/10.1145/3173574.3173998}
\showDOI{\tempurl}


\bibitem[Kalnikaitė and Whittaker(2011)]%
        {KALNIKAITE2011298}
\bibfield{author}{\bibinfo{person}{Vaiva Kalnikaitė} {and} \bibinfo{person}{Steve Whittaker}.} \bibinfo{year}{2011}\natexlab{}.
\newblock \showarticletitle{A saunter down memory lane: Digital reflection on personal mementos}.
\newblock \bibinfo{journal}{\emph{International Journal of Human-Computer Studies}} \bibinfo{volume}{69}, \bibinfo{number}{5} (\bibinfo{year}{2011}), \bibinfo{pages}{298--310}.
\newblock
\showISSN{1071-5819}
\urldef\tempurl%
\url{https://doi.org/10.1016/j.ijhcs.2010.12.004}
\showDOI{\tempurl}
\newblock
\shownote{Designing for Reflection on Personal Experience}.


\bibitem[Kang et~al\mbox{.}(2022)]%
        {10.1145/3526114.3558722}
\bibfield{author}{\bibinfo{person}{Yixiao Kang}, \bibinfo{person}{Zhenglin Zhang}, \bibinfo{person}{Meiqi Zhao}, \bibinfo{person}{Xuanhui Yang}, {and} \bibinfo{person}{Xubo Yang}.} \bibinfo{year}{2022}\natexlab{}.
\newblock \showarticletitle{Tie Memories to E-souvenirs: Hybrid Tangible AR Souvenirs in the Museum}. In \bibinfo{booktitle}{\emph{Adjunct Proceedings of the 35th Annual ACM Symposium on User Interface Software and Technology}} (Bend, OR, USA) \emph{(\bibinfo{series}{UIST '22 Adjunct})}. \bibinfo{publisher}{Association for Computing Machinery}, \bibinfo{address}{New York, NY, USA}, Article \bibinfo{articleno}{33}, \bibinfo{numpages}{3}~pages.
\newblock
\showISBNx{9781450393218}
\urldef\tempurl%
\url{https://doi.org/10.1145/3526114.3558722}
\showDOI{\tempurl}


\bibitem[Kawamura et~al\mbox{.}(2007)]%
        {kawamura2007ubiquitous}
\bibfield{author}{\bibinfo{person}{Tatsuyuki Kawamura}, \bibinfo{person}{Tomohiro Fukuhara}, \bibinfo{person}{Hideaki Takeda}, \bibinfo{person}{Yasuyuki Kono}, {and} \bibinfo{person}{Masatsugu Kidode}.} \bibinfo{year}{2007}\natexlab{}.
\newblock \showarticletitle{Ubiquitous memories: a memory externalization system using physical objects}.
\newblock \bibinfo{journal}{\emph{Personal and Ubiquitous Computing}}  \bibinfo{volume}{11} (\bibinfo{year}{2007}), \bibinfo{pages}{287--298}.
\newblock


\bibitem[Kim and Bianchi(2022)]%
        {kim2022spinocchietto}
\bibfield{author}{\bibinfo{person}{Myung~Jin Kim} {and} \bibinfo{person}{Andrea Bianchi}.} \bibinfo{year}{2022}\natexlab{}.
\newblock \showarticletitle{SpinOcchietto: A Wearable Skin-Slip Haptic Device for Rendering Width and Motion of Objects Gripped Between the Fingertips}. In \bibinfo{booktitle}{\emph{Adjunct Proceedings of the 35th Annual ACM Symposium on User Interface Software and Technology}}. \bibinfo{pages}{1--3}.
\newblock


\bibitem[Kirk and Sellen(2010a)]%
        {10.1145/1806923.1806924}
\bibfield{author}{\bibinfo{person}{David~S. Kirk} {and} \bibinfo{person}{Abigail Sellen}.} \bibinfo{year}{2010}\natexlab{a}.
\newblock \showarticletitle{On Human Remains: Values and Practice in the Home Archiving of Cherished Objects}.
\newblock \bibinfo{journal}{\emph{ACM Trans. Comput.-Hum. Interact.}} \bibinfo{volume}{17}, \bibinfo{number}{3}, Article \bibinfo{articleno}{10} (\bibinfo{date}{jul} \bibinfo{year}{2010}), \bibinfo{numpages}{43}~pages.
\newblock
\showISSN{1073-0516}
\urldef\tempurl%
\url{https://doi.org/10.1145/1806923.1806924}
\showDOI{\tempurl}


\bibitem[Kirk and Sellen(2010b)]%
        {kirk2010human}
\bibfield{author}{\bibinfo{person}{David~S Kirk} {and} \bibinfo{person}{Abigail Sellen}.} \bibinfo{year}{2010}\natexlab{b}.
\newblock \showarticletitle{On human remains: Values and practice in the home archiving of cherished objects}.
\newblock \bibinfo{journal}{\emph{ACM Transactions on Computer-Human Interaction (TOCHI)}} \bibinfo{volume}{17}, \bibinfo{number}{3} (\bibinfo{year}{2010}), \bibinfo{pages}{1--43}.
\newblock


\bibitem[Kleinberger et~al\mbox{.}(2019)]%
        {10.1145/3332165.3347877}
\bibfield{author}{\bibinfo{person}{Rebecca Kleinberger}, \bibinfo{person}{Alexandra Rieger}, \bibinfo{person}{Janelle Sands}, {and} \bibinfo{person}{Janet Baker}.} \bibinfo{year}{2019}\natexlab{}.
\newblock \showarticletitle{Supporting Elder Connectedness through Cognitively Sustainable Design Interactions with the Memory Music Box}. In \bibinfo{booktitle}{\emph{Proceedings of the 32nd Annual ACM Symposium on User Interface Software and Technology}} (New Orleans, LA, USA) \emph{(\bibinfo{series}{UIST '19})}. \bibinfo{publisher}{Association for Computing Machinery}, \bibinfo{address}{New York, NY, USA}, \bibinfo{pages}{355–369}.
\newblock
\showISBNx{9781450368162}
\urldef\tempurl%
\url{https://doi.org/10.1145/3332165.3347877}
\showDOI{\tempurl}


\bibitem[Krokos et~al\mbox{.}(2019)]%
        {krokos2019virtual}
\bibfield{author}{\bibinfo{person}{Eric Krokos}, \bibinfo{person}{Catherine Plaisant}, {and} \bibinfo{person}{Amitabh Varshney}.} \bibinfo{year}{2019}\natexlab{}.
\newblock \showarticletitle{Virtual memory palaces: immersion aids recall}.
\newblock \bibinfo{journal}{\emph{Virtual reality}} \bibinfo{volume}{23}, \bibinfo{number}{1} (\bibinfo{year}{2019}), \bibinfo{pages}{1--15}.
\newblock


\bibitem[Latoschik et~al\mbox{.}(2017)]%
        {10.1145/3139131.3139156}
\bibfield{author}{\bibinfo{person}{Marc~Erich Latoschik}, \bibinfo{person}{Daniel Roth}, \bibinfo{person}{Dominik Gall}, \bibinfo{person}{Jascha Achenbach}, \bibinfo{person}{Thomas Waltemate}, {and} \bibinfo{person}{Mario Botsch}.} \bibinfo{year}{2017}\natexlab{}.
\newblock \showarticletitle{The effect of avatar realism in immersive social virtual realities}. In \bibinfo{booktitle}{\emph{Proceedings of the 23rd ACM Symposium on Virtual Reality Software and Technology}} (Gothenburg, Sweden) \emph{(\bibinfo{series}{VRST '17})}. \bibinfo{publisher}{Association for Computing Machinery}, \bibinfo{address}{New York, NY, USA}, Article \bibinfo{articleno}{39}, \bibinfo{numpages}{10}~pages.
\newblock
\showISBNx{9781450355483}
\urldef\tempurl%
\url{https://doi.org/10.1145/3139131.3139156}
\showDOI{\tempurl}


\bibitem[Lekan(2016)]%
        {lekan2016virtual}
\bibfield{author}{\bibinfo{person}{Stephen Lekan}.} \bibinfo{year}{2016}\natexlab{}.
\newblock \emph{\bibinfo{title}{Virtual Environments, Rendered Realism and Their Effects on Spatial Memory}}.
\newblock \bibinfo{thesistype}{Ph.\,D. Dissertation}.
\newblock


\bibitem[Lesser et~al\mbox{.}(2022)]%
        {lesser2022loki}
\bibfield{author}{\bibinfo{person}{Steve Lesser}, \bibinfo{person}{Alexey Stomakhin}, \bibinfo{person}{Gilles Daviet}, \bibinfo{person}{Joel Wretborn}, \bibinfo{person}{John Edholm}, \bibinfo{person}{Noh-Hoon Lee}, \bibinfo{person}{Eston Schweickart}, \bibinfo{person}{Xiao Zhai}, \bibinfo{person}{Sean Flynn}, {and} \bibinfo{person}{Andrew Moffat}.} \bibinfo{year}{2022}\natexlab{}.
\newblock \showarticletitle{Loki: a unified multiphysics simulation framework for production}.
\newblock \bibinfo{journal}{\emph{ACM Transactions on Graphics (TOG)}} \bibinfo{volume}{41}, \bibinfo{number}{4} (\bibinfo{year}{2022}), \bibinfo{pages}{1--20}.
\newblock


\bibitem[Li et~al\mbox{.}(2020)]%
        {li2020facilitating}
\bibfield{author}{\bibinfo{person}{Cun Li}, \bibinfo{person}{Jun Hu}, \bibinfo{person}{Bart Hengeveld}, {and} \bibinfo{person}{Caroline Hummels}.} \bibinfo{year}{2020}\natexlab{}.
\newblock \showarticletitle{Facilitating storytelling and preservation of mementos for the elderly through tangible interface}. In \bibinfo{booktitle}{\emph{Human Systems Engineering and Design II: Proceedings of the 2nd International Conference on Human Systems Engineering and Design (IHSED2019): Future Trends and Applications, September 16-18, 2019, Universit{\"a}t der Bundeswehr M{\"u}nchen, Munich, Germany}}. Springer, \bibinfo{pages}{508--514}.
\newblock


\bibitem[Li et~al\mbox{.}(2023b)]%
        {li2023pacnerfphysicsaugmentedcontinuum}
\bibfield{author}{\bibinfo{person}{Xuan Li}, \bibinfo{person}{Yi-Ling Qiao}, \bibinfo{person}{Peter~Yichen Chen}, \bibinfo{person}{Krishna~Murthy Jatavallabhula}, \bibinfo{person}{Ming Lin}, \bibinfo{person}{Chenfanfu Jiang}, {and} \bibinfo{person}{Chuang Gan}.} \bibinfo{year}{2023}\natexlab{b}.
\newblock \bibinfo{title}{PAC-NeRF: Physics Augmented Continuum Neural Radiance Fields for Geometry-Agnostic System Identification}.
\newblock
\newblock
\showeprint[arxiv]{2303.05512}~[cs.CV]
\urldef\tempurl%
\url{https://arxiv.org/abs/2303.05512}
\showURL{%
\tempurl}


\bibitem[Li et~al\mbox{.}(2023a)]%
        {10.1145/3610903}
\bibfield{author}{\bibinfo{person}{Zisu Li}, \bibinfo{person}{Li Feng}, \bibinfo{person}{Chen Liang}, \bibinfo{person}{Yuru Huang}, {and} \bibinfo{person}{Mingming Fan}.} \bibinfo{year}{2023}\natexlab{a}.
\newblock \showarticletitle{Exploring the Opportunities of AR for Enriching Storytelling with Family Photos between Grandparents and Grandchildren}.
\newblock \bibinfo{journal}{\emph{Proc. ACM Interact. Mob. Wearable Ubiquitous Technol.}} \bibinfo{volume}{7}, \bibinfo{number}{3}, Article \bibinfo{articleno}{108} (\bibinfo{date}{sep} \bibinfo{year}{2023}), \bibinfo{numpages}{26}~pages.
\newblock
\urldef\tempurl%
\url{https://doi.org/10.1145/3610903}
\showDOI{\tempurl}


\bibitem[Liao et~al\mbox{.}(2022)]%
        {liao2022realitytalk}
\bibfield{author}{\bibinfo{person}{Jian Liao}, \bibinfo{person}{Adnan Karim}, \bibinfo{person}{Shivesh~Singh Jadon}, \bibinfo{person}{Rubaiat~Habib Kazi}, {and} \bibinfo{person}{Ryo Suzuki}.} \bibinfo{year}{2022}\natexlab{}.
\newblock \showarticletitle{RealityTalk: Real-time speech-driven augmented presentation for AR live storytelling}. In \bibinfo{booktitle}{\emph{Proceedings of the 35th Annual ACM Symposium on User Interface Software and Technology}}. \bibinfo{pages}{1--12}.
\newblock


\bibitem[Liu et~al\mbox{.}(2023)]%
        {10.1145/3544549.3585588}
\bibfield{author}{\bibinfo{person}{Yimeng Liu}, \bibinfo{person}{Jacob Ritchie}, \bibinfo{person}{Sven Kratz}, \bibinfo{person}{Misha Sra}, \bibinfo{person}{Brian~A. Smith}, \bibinfo{person}{Andr\'{e}s Monroy-Hern\'{a}ndez}, {and} \bibinfo{person}{Rajan Vaish}.} \bibinfo{year}{2023}\natexlab{}.
\newblock \showarticletitle{Memento Player: Shared Multi-Perspective Playback of Volumetrically-Captured Moments in Augmented Reality}. In \bibinfo{booktitle}{\emph{Extended Abstracts of the 2023 CHI Conference on Human Factors in Computing Systems}} (Hamburg, Germany) \emph{(\bibinfo{series}{CHI EA '23})}. \bibinfo{publisher}{Association for Computing Machinery}, \bibinfo{address}{New York, NY, USA}, Article \bibinfo{articleno}{205}, \bibinfo{numpages}{9}~pages.
\newblock
\showISBNx{9781450394222}
\urldef\tempurl%
\url{https://doi.org/10.1145/3544549.3585588}
\showDOI{\tempurl}


\bibitem[Liu et~al\mbox{.}(2024)]%
        {liu2024sora}
\bibfield{author}{\bibinfo{person}{Yixin Liu}, \bibinfo{person}{Kai Zhang}, \bibinfo{person}{Yuan Li}, \bibinfo{person}{Zhiling Yan}, \bibinfo{person}{Chujie Gao}, \bibinfo{person}{Ruoxi Chen}, \bibinfo{person}{Zhengqing Yuan}, \bibinfo{person}{Yue Huang}, \bibinfo{person}{Hanchi Sun}, \bibinfo{person}{Jianfeng Gao}, {et~al\mbox{.}}} \bibinfo{year}{2024}\natexlab{}.
\newblock \showarticletitle{Sora: A Review on Background, Technology, Limitations, and Opportunities of Large Vision Models}.
\newblock \bibinfo{journal}{\emph{arXiv preprint arXiv:2402.17177}} (\bibinfo{year}{2024}).
\newblock


\bibitem[L{\"u} and Li(2013)]%
        {lu2013gesture}
\bibfield{author}{\bibinfo{person}{Hao L{\"u}} {and} \bibinfo{person}{Yang Li}.} \bibinfo{year}{2013}\natexlab{}.
\newblock \showarticletitle{Gesture studio: authoring multi-touch interactions through demonstration and declaration}. In \bibinfo{booktitle}{\emph{Proceedings of the SIGCHI Conference on Human Factors in Computing Systems}}. \bibinfo{pages}{257--266}.
\newblock


\bibitem[Maddali and Lazar(2023)]%
        {maddali2023understanding}
\bibfield{author}{\bibinfo{person}{Hanuma~Teja Maddali} {and} \bibinfo{person}{Amanda Lazar}.} \bibinfo{year}{2023}\natexlab{}.
\newblock \showarticletitle{Understanding context to capture when reconstructing meaningful spaces for remote instruction and connecting in XR}. In \bibinfo{booktitle}{\emph{Proceedings of the 2023 CHI conference on Human Factors in Computing Systems}}. \bibinfo{pages}{1--18}.
\newblock


\bibitem[Marschall(2019)]%
        {marschall2019memory}
\bibfield{author}{\bibinfo{person}{Sabine Marschall}.} \bibinfo{year}{2019}\natexlab{}.
\newblock \showarticletitle{‘Memory objects’: Material objects and memories of home in the context of intra-African mobility}.
\newblock \bibinfo{journal}{\emph{Journal of Material Culture}} \bibinfo{volume}{24}, \bibinfo{number}{3} (\bibinfo{year}{2019}), \bibinfo{pages}{253--269}.
\newblock


\bibitem[Mayer and Rauber(2009)]%
        {mayer2009establishing}
\bibfield{author}{\bibinfo{person}{Rudolf Mayer} {and} \bibinfo{person}{Andreas Rauber}.} \bibinfo{year}{2009}\natexlab{}.
\newblock \showarticletitle{Establishing context of digital objects’ creation, content and usage}. In \bibinfo{booktitle}{\emph{Proc. Int. Workshop on Innovation in Digital Preservation}}.
\newblock


\bibitem[McMahan et~al\mbox{.}(2016)]%
        {mcmahan2016interaction}
\bibfield{author}{\bibinfo{person}{Ryan~P McMahan}, \bibinfo{person}{Chengyuan Lai}, {and} \bibinfo{person}{Swaroop~K Pal}.} \bibinfo{year}{2016}\natexlab{}.
\newblock \showarticletitle{Interaction fidelity: the uncanny valley of virtual reality interactions}. In \bibinfo{booktitle}{\emph{Virtual, Augmented and Mixed Reality: 8th International Conference, VAMR 2016, Held as Part of HCI International 2016, Toronto, Canada, July 17-22, 2016. Proceedings 8}}. Springer, \bibinfo{pages}{59--70}.
\newblock


\bibitem[Mezaris et~al\mbox{.}(2018)]%
        {mezaris2018personal}
\bibfield{author}{\bibinfo{person}{Vasileios Mezaris}, \bibinfo{person}{Claudia Nieder{\'e}e}, {and} \bibinfo{person}{Robert~H Logie}.} \bibinfo{year}{2018}\natexlab{}.
\newblock \bibinfo{booktitle}{\emph{Personal Multimedia Preservation: Remembering or forgetting images and video}}.
\newblock \bibinfo{publisher}{Springer}.
\newblock


\bibitem[Millington(2010)]%
        {millington2010game}
\bibfield{author}{\bibinfo{person}{Ian Millington}.} \bibinfo{year}{2010}\natexlab{}.
\newblock \bibinfo{booktitle}{\emph{Game physics engine development: how to build a robust commercial-grade physics engine for your game}}.
\newblock \bibinfo{publisher}{CRC Press}.
\newblock


\bibitem[Neumann et~al\mbox{.}(2017)]%
        {10.1145/3027063.3052756}
\bibfield{author}{\bibinfo{person}{Stephanie Neumann}, \bibinfo{person}{Richard Banks}, {and} \bibinfo{person}{Marian D\"{o}rk}.} \bibinfo{year}{2017}\natexlab{}.
\newblock \showarticletitle{Memory Dialogue: Exploring Artefact-Based Memory Sharing}. In \bibinfo{booktitle}{\emph{Proceedings of the 2017 CHI Conference Extended Abstracts on Human Factors in Computing Systems}} (Denver, Colorado, USA) \emph{(\bibinfo{series}{CHI EA '17})}. \bibinfo{publisher}{Association for Computing Machinery}, \bibinfo{address}{New York, NY, USA}, \bibinfo{pages}{884–895}.
\newblock
\showISBNx{9781450346566}
\urldef\tempurl%
\url{https://doi.org/10.1145/3027063.3052756}
\showDOI{\tempurl}


\bibitem[Neumayer et~al\mbox{.}(2005)]%
        {neumayer2005content}
\bibfield{author}{\bibinfo{person}{Robert Neumayer}, \bibinfo{person}{Thomas Lidy}, {and} \bibinfo{person}{Andreas Rauber}.} \bibinfo{year}{2005}\natexlab{}.
\newblock \bibinfo{booktitle}{\emph{Content-based organization of digital audio collections}}.
\newblock \bibinfo{publisher}{na}.
\newblock


\bibitem[Nunes et~al\mbox{.}(2008)]%
        {10.1145/1394445.1394472}
\bibfield{author}{\bibinfo{person}{Michael Nunes}, \bibinfo{person}{Saul Greenberg}, {and} \bibinfo{person}{Carman Neustaedter}.} \bibinfo{year}{2008}\natexlab{}.
\newblock \showarticletitle{Sharing Digital Photographs in the Home through Physical Mementos, Souvenirs, and Keepsakes}. In \bibinfo{booktitle}{\emph{Proceedings of the 7th ACM Conference on Designing Interactive Systems}} (Cape Town, South Africa) \emph{(\bibinfo{series}{DIS '08})}. \bibinfo{publisher}{Association for Computing Machinery}, \bibinfo{address}{New York, NY, USA}, \bibinfo{pages}{250–260}.
\newblock
\showISBNx{9781605580029}
\urldef\tempurl%
\url{https://doi.org/10.1145/1394445.1394472}
\showDOI{\tempurl}


\bibitem[Nunes et~al\mbox{.}(2009)]%
        {nunes2009using}
\bibfield{author}{\bibinfo{person}{Michael Nunes}, \bibinfo{person}{Saul Greenberg}, {and} \bibinfo{person}{Carman Neustaedter}.} \bibinfo{year}{2009}\natexlab{}.
\newblock \showarticletitle{Using physical memorabilia as opportunities to move into collocated digital photo-sharing}.
\newblock \bibinfo{journal}{\emph{International Journal of Human-Computer Studies}} \bibinfo{volume}{67}, \bibinfo{number}{12} (\bibinfo{year}{2009}), \bibinfo{pages}{1087--1111}.
\newblock


\bibitem[Odom et~al\mbox{.}(2012)]%
        {10.1145/2317956.2318055}
\bibfield{author}{\bibinfo{person}{William Odom}, \bibinfo{person}{Mark Selby}, \bibinfo{person}{Abigail Sellen}, \bibinfo{person}{David Kirk}, \bibinfo{person}{Richard Banks}, {and} \bibinfo{person}{Tim Regan}.} \bibinfo{year}{2012}\natexlab{}.
\newblock \showarticletitle{Photobox: on the design of a slow technology}. In \bibinfo{booktitle}{\emph{Proceedings of the Designing Interactive Systems Conference}} (Newcastle Upon Tyne, United Kingdom) \emph{(\bibinfo{series}{DIS '12})}. \bibinfo{publisher}{Association for Computing Machinery}, \bibinfo{address}{New York, NY, USA}, \bibinfo{pages}{665–668}.
\newblock
\showISBNx{9781450312103}
\urldef\tempurl%
\url{https://doi.org/10.1145/2317956.2318055}
\showDOI{\tempurl}


\bibitem[Peesapati et~al\mbox{.}(2010)]%
        {10.1145/1753326.1753635}
\bibfield{author}{\bibinfo{person}{S.~Tejaswi Peesapati}, \bibinfo{person}{Victoria Schwanda}, \bibinfo{person}{Johnathon Schultz}, \bibinfo{person}{Matt Lepage}, \bibinfo{person}{So-yae Jeong}, {and} \bibinfo{person}{Dan Cosley}.} \bibinfo{year}{2010}\natexlab{}.
\newblock \showarticletitle{Pensieve: supporting everyday reminiscence}. In \bibinfo{booktitle}{\emph{Proceedings of the SIGCHI Conference on Human Factors in Computing Systems}} (Atlanta, Georgia, USA) \emph{(\bibinfo{series}{CHI '10})}. \bibinfo{publisher}{Association for Computing Machinery}, \bibinfo{address}{New York, NY, USA}, \bibinfo{pages}{2027–2036}.
\newblock
\showISBNx{9781605589299}
\urldef\tempurl%
\url{https://doi.org/10.1145/1753326.1753635}
\showDOI{\tempurl}


\bibitem[Petrelli et~al\mbox{.}(2014)]%
        {petrelli2014photo}
\bibfield{author}{\bibinfo{person}{Daniela Petrelli}, \bibinfo{person}{Simon Bowen}, {and} \bibinfo{person}{Steve Whittaker}.} \bibinfo{year}{2014}\natexlab{}.
\newblock \showarticletitle{Photo mementos: Designing digital media to represent ourselves at home}.
\newblock \bibinfo{journal}{\emph{International Journal of Human-Computer Studies}} \bibinfo{volume}{72}, \bibinfo{number}{3} (\bibinfo{year}{2014}), \bibinfo{pages}{320--336}.
\newblock


\bibitem[Petrelli et~al\mbox{.}(2009)]%
        {10.1145/1518701.1518966}
\bibfield{author}{\bibinfo{person}{Daniela Petrelli}, \bibinfo{person}{Elise van~den Hoven}, {and} \bibinfo{person}{Steve Whittaker}.} \bibinfo{year}{2009}\natexlab{}.
\newblock \showarticletitle{Making History: Intentional Capture of Future Memories}. In \bibinfo{booktitle}{\emph{Proceedings of the SIGCHI Conference on Human Factors in Computing Systems}} (Boston, MA, USA) \emph{(\bibinfo{series}{CHI '09})}. \bibinfo{publisher}{Association for Computing Machinery}, \bibinfo{address}{New York, NY, USA}, \bibinfo{pages}{1723–1732}.
\newblock
\showISBNx{9781605582467}
\urldef\tempurl%
\url{https://doi.org/10.1145/1518701.1518966}
\showDOI{\tempurl}


\bibitem[Petrelli and Whittaker(2010)]%
        {petrelli2010family}
\bibfield{author}{\bibinfo{person}{Daniela Petrelli} {and} \bibinfo{person}{Steve Whittaker}.} \bibinfo{year}{2010}\natexlab{}.
\newblock \showarticletitle{Family memories in the home: contrasting physical and digital mementos}.
\newblock \bibinfo{journal}{\emph{Personal and Ubiquitous Computing}}  \bibinfo{volume}{14} (\bibinfo{year}{2010}), \bibinfo{pages}{153--169}.
\newblock


\bibitem[Petrelli et~al\mbox{.}(2008)]%
        {petrelli2008autotopography}
\bibfield{author}{\bibinfo{person}{Daniela Petrelli}, \bibinfo{person}{Steve Whittaker}, {and} \bibinfo{person}{Jens Brockmeier}.} \bibinfo{year}{2008}\natexlab{}.
\newblock \showarticletitle{AutoTopography: what can physical mementos tell us about digital memories?}. In \bibinfo{booktitle}{\emph{Proceedings of the SIGCHI conference on Human Factors in computing systems}}. \bibinfo{pages}{53--62}.
\newblock


\bibitem[Podell et~al\mbox{.}(2023)]%
        {podell2023sdxl}
\bibfield{author}{\bibinfo{person}{Dustin Podell}, \bibinfo{person}{Zion English}, \bibinfo{person}{Kyle Lacey}, \bibinfo{person}{Andreas Blattmann}, \bibinfo{person}{Tim Dockhorn}, \bibinfo{person}{Jonas M{\"u}ller}, \bibinfo{person}{Joe Penna}, {and} \bibinfo{person}{Robin Rombach}.} \bibinfo{year}{2023}\natexlab{}.
\newblock \showarticletitle{Sdxl: Improving latent diffusion models for high-resolution image synthesis}.
\newblock \bibinfo{journal}{\emph{arXiv preprint arXiv:2307.01952}} (\bibinfo{year}{2023}).
\newblock


\bibitem[Pollio and Foote(1971)]%
        {pollio1971memory}
\bibfield{author}{\bibinfo{person}{Howard~R Pollio} {and} \bibinfo{person}{Russell Foote}.} \bibinfo{year}{1971}\natexlab{}.
\newblock \showarticletitle{Memory as a reconstructive process}.
\newblock \bibinfo{journal}{\emph{British Journal of Psychology}} \bibinfo{volume}{62}, \bibinfo{number}{1} (\bibinfo{year}{1971}), \bibinfo{pages}{53--58}.
\newblock


\bibitem[Potter(2010)]%
        {potter2010embodied}
\bibfield{author}{\bibinfo{person}{John Potter}.} \bibinfo{year}{2010}\natexlab{}.
\newblock \showarticletitle{Embodied Memory and Curatorship in Children's Digital Video Production.}
\newblock \bibinfo{journal}{\emph{English Teaching: Practice and Critique}} \bibinfo{volume}{9}, \bibinfo{number}{1} (\bibinfo{year}{2010}), \bibinfo{pages}{22--35}.
\newblock


\bibitem[Ren et~al\mbox{.}(2024)]%
        {ren2024dreamgaussian4dgenerative4dgaussian}
\bibfield{author}{\bibinfo{person}{Jiawei Ren}, \bibinfo{person}{Liang Pan}, \bibinfo{person}{Jiaxiang Tang}, \bibinfo{person}{Chi Zhang}, \bibinfo{person}{Ang Cao}, \bibinfo{person}{Gang Zeng}, {and} \bibinfo{person}{Ziwei Liu}.} \bibinfo{year}{2024}\natexlab{}.
\newblock \bibinfo{title}{DreamGaussian4D: Generative 4D Gaussian Splatting}.
\newblock
\newblock
\showeprint[arxiv]{2312.17142}~[cs.CV]
\urldef\tempurl%
\url{https://arxiv.org/abs/2312.17142}
\showURL{%
\tempurl}


\bibitem[Runardotter(2009)]%
        {runardotter2009organizing}
\bibfield{author}{\bibinfo{person}{Mari Runardotter}.} \bibinfo{year}{2009}\natexlab{}.
\newblock \emph{\bibinfo{title}{On organizing for digital preservation}}.
\newblock \bibinfo{thesistype}{Ph.\,D. Dissertation}. \bibinfo{school}{Lule{\aa} tekniska universitet}.
\newblock


\bibitem[Sas et~al\mbox{.}(2015)]%
        {10.1145/2702613.2732842}
\bibfield{author}{\bibinfo{person}{Corina Sas}, \bibinfo{person}{Scott Challioner}, \bibinfo{person}{Christopher Clarke}, \bibinfo{person}{Ross Wilson}, \bibinfo{person}{Alina Coman}, \bibinfo{person}{Sarah Clinch}, \bibinfo{person}{Mike Harding}, {and} \bibinfo{person}{Nigel Davies}.} \bibinfo{year}{2015}\natexlab{}.
\newblock \showarticletitle{Self-Defining Memory Cues: Creative Expression and Emotional Meaning}. In \bibinfo{booktitle}{\emph{Proceedings of the 33rd Annual ACM Conference Extended Abstracts on Human Factors in Computing Systems}} (<conf-loc>, <city>Seoul</city>, <country>Republic of Korea</country>, </conf-loc>) \emph{(\bibinfo{series}{CHI EA '15})}. \bibinfo{publisher}{Association for Computing Machinery}, \bibinfo{address}{New York, NY, USA}, \bibinfo{pages}{2013–2018}.
\newblock
\showISBNx{9781450331463}
\urldef\tempurl%
\url{https://doi.org/10.1145/2702613.2732842}
\showDOI{\tempurl}


\bibitem[Scarfe and Glennerster(2015)]%
        {scarfe2015using}
\bibfield{author}{\bibinfo{person}{Peter Scarfe} {and} \bibinfo{person}{Andrew Glennerster}.} \bibinfo{year}{2015}\natexlab{}.
\newblock \showarticletitle{Using high-fidelity virtual reality to study perception in freely moving observers}.
\newblock \bibinfo{journal}{\emph{Journal of vision}} \bibinfo{volume}{15}, \bibinfo{number}{9} (\bibinfo{year}{2015}), \bibinfo{pages}{3--3}.
\newblock


\bibitem[Seymour et~al\mbox{.}(2021)]%
        {seymour2021have}
\bibfield{author}{\bibinfo{person}{Mike Seymour}, \bibinfo{person}{Lingyao~Ivy Yuan}, \bibinfo{person}{Alan Dennis}, \bibinfo{person}{Kai Riemer}, {et~al\mbox{.}}} \bibinfo{year}{2021}\natexlab{}.
\newblock \showarticletitle{Have we crossed the uncanny valley? Understanding affinity, trustworthiness, and preference for realistic digital humans in immersive environments}.
\newblock \bibinfo{journal}{\emph{Journal of the Association for Information Systems}} \bibinfo{volume}{22}, \bibinfo{number}{3} (\bibinfo{year}{2021}), \bibinfo{pages}{9}.
\newblock


\bibitem[Somerville and EchoHawk(2011)]%
        {somerville2011recuerdos}
\bibfield{author}{\bibinfo{person}{Mary~M Somerville} {and} \bibinfo{person}{Dana EchoHawk}.} \bibinfo{year}{2011}\natexlab{}.
\newblock \showarticletitle{Recuerdos hablados/memories spoken: Toward the co-creation of digital knowledge with community significance}.
\newblock \bibinfo{journal}{\emph{Library Trends}} \bibinfo{volume}{59}, \bibinfo{number}{4} (\bibinfo{year}{2011}), \bibinfo{pages}{650--662}.
\newblock


\bibitem[Song et~al\mbox{.}(2023)]%
        {song2023consistency}
\bibfield{author}{\bibinfo{person}{Yang Song}, \bibinfo{person}{Prafulla Dhariwal}, \bibinfo{person}{Mark Chen}, {and} \bibinfo{person}{Ilya Sutskever}.} \bibinfo{year}{2023}\natexlab{}.
\newblock \showarticletitle{Consistency models}.
\newblock \bibinfo{journal}{\emph{arXiv preprint arXiv:2303.01469}} (\bibinfo{year}{2023}).
\newblock


\bibitem[Stemasov et~al\mbox{.}(2020)]%
        {stemasov2020mix}
\bibfield{author}{\bibinfo{person}{Evgeny Stemasov}, \bibinfo{person}{Tobias Wagner}, \bibinfo{person}{Jan Gugenheimer}, {and} \bibinfo{person}{Enrico Rukzio}.} \bibinfo{year}{2020}\natexlab{}.
\newblock \showarticletitle{Mix\&Match: Towards omitting modelling through in-situ remixing of model repository artifacts in mixed reality}. In \bibinfo{booktitle}{\emph{Proceedings of the 2020 CHI Conference on Human Factors in Computing Systems}}. \bibinfo{pages}{1--12}.
\newblock


\bibitem[Tan and Rahaman(2009)]%
        {tan2009virtual}
\bibfield{author}{\bibinfo{person}{Beng-Kiang Tan} {and} \bibinfo{person}{Hafizur Rahaman}.} \bibinfo{year}{2009}\natexlab{}.
\newblock \showarticletitle{Virtual heritage: Reality and criticism}. In \bibinfo{booktitle}{\emph{CAAD futures}}. Les Presses de l'Universit{\'e} de Montr{\'e}al, \bibinfo{pages}{143--156}.
\newblock


\bibitem[Tan et~al\mbox{.}(2018)]%
        {10.1145/3287069}
\bibfield{author}{\bibinfo{person}{Neille-Ann~H. Tan}, \bibinfo{person}{Han Sha}, \bibinfo{person}{Eda Celen}, \bibinfo{person}{Phucanh Tran}, \bibinfo{person}{Kelly Wang}, \bibinfo{person}{Gifford Cheung}, \bibinfo{person}{Philip Hinch}, {and} \bibinfo{person}{Jeff Huang}.} \bibinfo{year}{2018}\natexlab{}.
\newblock \showarticletitle{Rewind: Automatically Reconstructing Everyday Memories with First-Person Perspectives}.
\newblock \bibinfo{journal}{\emph{Proc. ACM Interact. Mob. Wearable Ubiquitous Technol.}} \bibinfo{volume}{2}, \bibinfo{number}{4}, Article \bibinfo{articleno}{191} (\bibinfo{date}{dec} \bibinfo{year}{2018}), \bibinfo{numpages}{20}~pages.
\newblock
\urldef\tempurl%
\url{https://doi.org/10.1145/3287069}
\showDOI{\tempurl}


\bibitem[Tatzgern et~al\mbox{.}(2022)]%
        {tatzgern2022airres}
\bibfield{author}{\bibinfo{person}{Markus Tatzgern}, \bibinfo{person}{Michael Domhardt}, \bibinfo{person}{Martin Wolf}, \bibinfo{person}{Michael Cenger}, \bibinfo{person}{Gerlinde Emsenhuber}, \bibinfo{person}{Radomir Dinic}, \bibinfo{person}{Nathalie Gerner}, {and} \bibinfo{person}{Arnulf Hartl}.} \bibinfo{year}{2022}\natexlab{}.
\newblock \showarticletitle{Airres mask: A precise and robust virtual reality breathing interface utilizing breathing resistance as output modality}. In \bibinfo{booktitle}{\emph{Proceedings of the 2022 CHI Conference on Human Factors in Computing Systems}}. \bibinfo{pages}{1--14}.
\newblock


\bibitem[Tsai et~al\mbox{.}(2014)]%
        {10.1145/2598510.2598589}
\bibfield{author}{\bibinfo{person}{Wenn-Chieh Tsai}, \bibinfo{person}{Po-Hao Wang}, \bibinfo{person}{Hung-Chi Lee}, \bibinfo{person}{Rung-Huei Liang}, {and} \bibinfo{person}{Jane Hsu}.} \bibinfo{year}{2014}\natexlab{}.
\newblock \showarticletitle{The reflexive printer: toward making sense of perceived drawbacks in technology-mediated reminiscence}. In \bibinfo{booktitle}{\emph{Proceedings of the 2014 Conference on Designing Interactive Systems}} (Vancouver, BC, Canada) \emph{(\bibinfo{series}{DIS '14})}. \bibinfo{publisher}{Association for Computing Machinery}, \bibinfo{address}{New York, NY, USA}, \bibinfo{pages}{995–1004}.
\newblock
\showISBNx{9781450329026}
\urldef\tempurl%
\url{https://doi.org/10.1145/2598510.2598589}
\showDOI{\tempurl}


\bibitem[Valtolina et~al\mbox{.}(2005)]%
        {valtolina2005dissemination}
\bibfield{author}{\bibinfo{person}{Stefano Valtolina}, \bibinfo{person}{Stefano Franzoni}, \bibinfo{person}{Pietro Mazzoleni}, {and} \bibinfo{person}{Elisa Bertino}.} \bibinfo{year}{2005}\natexlab{}.
\newblock \showarticletitle{Dissemination of cultural heritage content through virtual reality and multimedia techniques: A case study}. In \bibinfo{booktitle}{\emph{11th international multimedia modelling conference}}. IEEE, \bibinfo{pages}{214--221}.
\newblock


\bibitem[Van~Someren et~al\mbox{.}(1994)]%
        {van1994think}
\bibfield{author}{\bibinfo{person}{Maarten Van~Someren}, \bibinfo{person}{Yvonne~F Barnard}, {and} \bibinfo{person}{J Sandberg}.} \bibinfo{year}{1994}\natexlab{}.
\newblock \showarticletitle{The think aloud method: a practical approach to modelling cognitive}.
\newblock \bibinfo{journal}{\emph{London: AcademicPress}}  \bibinfo{volume}{11} (\bibinfo{year}{1994}), \bibinfo{pages}{29--41}.
\newblock


\bibitem[Vecchio et~al\mbox{.}(2024a)]%
        {10.1145/3688830}
\bibfield{author}{\bibinfo{person}{Giuseppe Vecchio}, \bibinfo{person}{Rosalie Martin}, \bibinfo{person}{Arthur Roullier}, \bibinfo{person}{Adrien Kaiser}, \bibinfo{person}{Romain Rouffet}, \bibinfo{person}{Valentin Deschaintre}, {and} \bibinfo{person}{Tamy Boubekeur}.} \bibinfo{year}{2024}\natexlab{a}.
\newblock \showarticletitle{ControlMat: A Controlled Generative Approach to Material Capture}.
\newblock \bibinfo{journal}{\emph{ACM Trans. Graph.}} \bibinfo{volume}{43}, \bibinfo{number}{5}, Article \bibinfo{articleno}{164} (\bibinfo{date}{Sept.} \bibinfo{year}{2024}), \bibinfo{numpages}{17}~pages.
\newblock
\showISSN{0730-0301}
\urldef\tempurl%
\url{https://doi.org/10.1145/3688830}
\showDOI{\tempurl}


\bibitem[Vecchio et~al\mbox{.}(2024b)]%
        {Vecchio_2024}
\bibfield{author}{\bibinfo{person}{Giuseppe Vecchio}, \bibinfo{person}{Renato Sortino}, \bibinfo{person}{Simone Palazzo}, {and} \bibinfo{person}{Concetto Spampinato}.} \bibinfo{year}{2024}\natexlab{b}.
\newblock \showarticletitle{MatFuse: Controllable Material Generation with Diffusion Models}. In \bibinfo{booktitle}{\emph{2024 IEEE/CVF Conference on Computer Vision and Pattern Recognition (CVPR)}}. \bibinfo{publisher}{IEEE}, \bibinfo{pages}{4429–4438}.
\newblock
\urldef\tempurl%
\url{https://doi.org/10.1109/cvpr52733.2024.00424}
\showDOI{\tempurl}


\bibitem[Wang et~al\mbox{.}(2023)]%
        {wang2023dreamvideo}
\bibfield{author}{\bibinfo{person}{Cong Wang}, \bibinfo{person}{Jiaxi Gu}, \bibinfo{person}{Panwen Hu}, \bibinfo{person}{Songcen Xu}, \bibinfo{person}{Hang Xu}, {and} \bibinfo{person}{Xiaodan Liang}.} \bibinfo{year}{2023}\natexlab{}.
\newblock \showarticletitle{DreamVideo: High-Fidelity Image-to-Video Generation with Image Retention and Text Guidance}.
\newblock \bibinfo{journal}{\emph{arXiv preprint arXiv:2312.03018}} (\bibinfo{year}{2023}).
\newblock


\bibitem[Wang et~al\mbox{.}(2021)]%
        {10.1145/3472749.3474769}
\bibfield{author}{\bibinfo{person}{Tianyi Wang}, \bibinfo{person}{Xun Qian}, \bibinfo{person}{Fengming He}, \bibinfo{person}{Xiyun Hu}, \bibinfo{person}{Yuanzhi Cao}, {and} \bibinfo{person}{Karthik Ramani}.} \bibinfo{year}{2021}\natexlab{}.
\newblock \showarticletitle{GesturAR: An Authoring System for Creating Freehand Interactive Augmented Reality Applications}. In \bibinfo{booktitle}{\emph{The 34th Annual ACM Symposium on User Interface Software and Technology}} (Virtual Event, USA) \emph{(\bibinfo{series}{UIST '21})}. \bibinfo{publisher}{Association for Computing Machinery}, \bibinfo{address}{New York, NY, USA}, \bibinfo{pages}{552–567}.
\newblock
\showISBNx{9781450386357}
\urldef\tempurl%
\url{https://doi.org/10.1145/3472749.3474769}
\showDOI{\tempurl}


\bibitem[West et~al\mbox{.}(2007)]%
        {west2007memento}
\bibfield{author}{\bibinfo{person}{David West}, \bibinfo{person}{Aaron Quigley}, {and} \bibinfo{person}{Judy Kay}.} \bibinfo{year}{2007}\natexlab{}.
\newblock \showarticletitle{MEMENTO: a digital-physical scrapbook for memory sharing}.
\newblock \bibinfo{journal}{\emph{Personal and Ubiquitous Computing}}  \bibinfo{volume}{11} (\bibinfo{year}{2007}), \bibinfo{pages}{313--328}.
\newblock


\bibitem[Whitlock et~al\mbox{.}(2020)]%
        {whitlock2020mrcat}
\bibfield{author}{\bibinfo{person}{Matt Whitlock}, \bibinfo{person}{Jake Mitchell}, \bibinfo{person}{Nick Pfeufer}, \bibinfo{person}{Brad Arnot}, \bibinfo{person}{Ryan Craig}, \bibinfo{person}{Bryce Wilson}, \bibinfo{person}{Brian Chung}, {and} \bibinfo{person}{Danielle~Albers Szafir}.} \bibinfo{year}{2020}\natexlab{}.
\newblock \showarticletitle{MRCAT: In situ prototyping of interactive AR environments}. In \bibinfo{booktitle}{\emph{International Conference on Human-Computer Interaction}}. Springer, \bibinfo{pages}{235--255}.
\newblock


\bibitem[Williamson and Brown(2008)]%
        {10.1145/1394445.1394461}
\bibfield{author}{\bibinfo{person}{John Williamson} {and} \bibinfo{person}{Lorna~M Brown}.} \bibinfo{year}{2008}\natexlab{}.
\newblock \showarticletitle{Flutter: directed random browsing of photo collections with a tangible interface}. In \bibinfo{booktitle}{\emph{Proceedings of the 7th ACM Conference on Designing Interactive Systems}} (Cape Town, South Africa) \emph{(\bibinfo{series}{DIS '08})}. \bibinfo{publisher}{Association for Computing Machinery}, \bibinfo{address}{New York, NY, USA}, \bibinfo{pages}{147–155}.
\newblock
\showISBNx{9781605580029}
\urldef\tempurl%
\url{https://doi.org/10.1145/1394445.1394461}
\showDOI{\tempurl}


\bibitem[Xie et~al\mbox{.}(2024)]%
        {xie2024physgaussian}
\bibfield{author}{\bibinfo{person}{Tianyi Xie}, \bibinfo{person}{Zeshun Zong}, \bibinfo{person}{Yuxing Qiu}, \bibinfo{person}{Xuan Li}, \bibinfo{person}{Yutao Feng}, \bibinfo{person}{Yin Yang}, {and} \bibinfo{person}{Chenfanfu Jiang}.} \bibinfo{year}{2024}\natexlab{}.
\newblock \bibinfo{title}{PhysGaussian: Physics-Integrated 3D Gaussians for Generative Dynamics}.
\newblock
\newblock
\showeprint[arxiv]{2311.12198}~[cs.GR]
\urldef\tempurl%
\url{https://arxiv.org/abs/2311.12198}
\showURL{%
\tempurl}


\bibitem[Yang et~al\mbox{.}(2006)]%
        {yang2006creating}
\bibfield{author}{\bibinfo{person}{Cheng Yang}, \bibinfo{person}{Dongmei Peng}, {and} \bibinfo{person}{Shouqian Sun}.} \bibinfo{year}{2006}\natexlab{}.
\newblock \showarticletitle{Creating a virtual activity for the intangible culture heritage}. In \bibinfo{booktitle}{\emph{16th International Conference on Artificial Reality and Telexistence--Workshops (ICAT'06)}}. IEEE, \bibinfo{pages}{636--641}.
\newblock


\bibitem[Zhang et~al\mbox{.}(2024)]%
        {zhang2024physdreamer}
\bibfield{author}{\bibinfo{person}{Tianyuan Zhang}, \bibinfo{person}{Hong-Xing Yu}, \bibinfo{person}{Rundi Wu}, \bibinfo{person}{Brandon~Y. Feng}, \bibinfo{person}{Changxi Zheng}, \bibinfo{person}{Noah Snavely}, \bibinfo{person}{Jiajun Wu}, {and} \bibinfo{person}{William~T. Freeman}.} \bibinfo{year}{2024}\natexlab{}.
\newblock \showarticletitle{{PhysDreamer}: Physics-Based Interaction with 3D Objects via Video Generation}.
\newblock \bibinfo{journal}{\emph{arxiv}} (\bibinfo{year}{2024}).
\newblock


\bibitem[Zhang et~al\mbox{.}(2023)]%
        {zhang2023amphion}
\bibfield{author}{\bibinfo{person}{Xueyao Zhang}, \bibinfo{person}{Liumeng Xue}, \bibinfo{person}{Yuancheng Wang}, \bibinfo{person}{Yicheng Gu}, \bibinfo{person}{Xi Chen}, \bibinfo{person}{Zihao Fang}, \bibinfo{person}{Haopeng Chen}, \bibinfo{person}{Lexiao Zou}, \bibinfo{person}{Chaoren Wang}, \bibinfo{person}{Jun Han}, {et~al\mbox{.}}} \bibinfo{year}{2023}\natexlab{}.
\newblock \showarticletitle{Amphion: An Open-Source Audio, Music and Speech Generation Toolkit}.
\newblock \bibinfo{journal}{\emph{arXiv preprint arXiv:2312.09911}} (\bibinfo{year}{2023}).
\newblock


\bibitem[Zhao et~al\mbox{.}(2024)]%
        {zhao2024automated3dphysicalsimulation}
\bibfield{author}{\bibinfo{person}{Haoyu Zhao}, \bibinfo{person}{Hao Wang}, \bibinfo{person}{Xingyue Zhao}, \bibinfo{person}{Hongqiu Wang}, \bibinfo{person}{Zhiyu Wu}, \bibinfo{person}{Chengjiang Long}, {and} \bibinfo{person}{Hua Zou}.} \bibinfo{year}{2024}\natexlab{}.
\newblock \bibinfo{title}{Automated 3D Physical Simulation of Open-world Scene with Gaussian Splatting}.
\newblock
\newblock
\showeprint[arxiv]{2411.12789}~[cs.CV]
\urldef\tempurl%
\url{https://arxiv.org/abs/2411.12789}
\showURL{%
\tempurl}


\bibitem[Zhong et~al\mbox{.}(2022)]%
        {10.1145/3532106.3533501}
\bibfield{author}{\bibinfo{person}{Ce Zhong}, \bibinfo{person}{Ron Wakkary}, \bibinfo{person}{William Odom}, \bibinfo{person}{Amy Yo~Sue Chen}, \bibinfo{person}{MinYoung Yoo}, {and} \bibinfo{person}{Doenja Oogjes}.} \bibinfo{year}{2022}\natexlab{}.
\newblock \showarticletitle{On the Design of deformTable: Attending to Temporality and Materiality for Supporting Everyday Interactions with a Shape-Changing Artifact}. In \bibinfo{booktitle}{\emph{Proceedings of the 2022 ACM Designing Interactive Systems Conference}} (<conf-loc>, <city>Virtual Event</city>, <country>Australia</country>, </conf-loc>) \emph{(\bibinfo{series}{DIS '22})}. \bibinfo{publisher}{Association for Computing Machinery}, \bibinfo{address}{New York, NY, USA}, \bibinfo{pages}{1555–1564}.
\newblock
\showISBNx{9781450393584}
\urldef\tempurl%
\url{https://doi.org/10.1145/3532106.3533501}
\showDOI{\tempurl}


\bibitem[Zhou et~al\mbox{.}(2022)]%
        {zhou2022tilegentileablecontrollablematerial}
\bibfield{author}{\bibinfo{person}{Xilong Zhou}, \bibinfo{person}{Miloš Hašan}, \bibinfo{person}{Valentin Deschaintre}, \bibinfo{person}{Paul Guerrero}, \bibinfo{person}{Kalyan Sunkavalli}, {and} \bibinfo{person}{Nima Kalantari}.} \bibinfo{year}{2022}\natexlab{}.
\newblock \bibinfo{title}{TileGen: Tileable, Controllable Material Generation and Capture}.
\newblock
\newblock
\showeprint[arxiv]{2206.05649}~[cs.GR]
\urldef\tempurl%
\url{https://arxiv.org/abs/2206.05649}
\showURL{%
\tempurl}


\bibitem[Zhu et~al\mbox{.}(2023)]%
        {zhu2023minigpt}
\bibfield{author}{\bibinfo{person}{Deyao Zhu}, \bibinfo{person}{Jun Chen}, \bibinfo{person}{Xiaoqian Shen}, \bibinfo{person}{Xiang Li}, {and} \bibinfo{person}{Mohamed Elhoseiny}.} \bibinfo{year}{2023}\natexlab{}.
\newblock \showarticletitle{Minigpt-4: Enhancing vision-language understanding with advanced large language models}.
\newblock \bibinfo{journal}{\emph{arXiv preprint arXiv:2304.10592}} (\bibinfo{year}{2023}).
\newblock


\end{thebibliography}

\appendix
\newpage
\onecolumn
\begin{table*}[tbh!]
  \centering
  \caption{\textbf{Descriptions of the pre-defined physical joint categories.}}
  ~\label{tab:joints}
    \vspace{-0.3cm}
    \resizebox{\linewidth}{!}{
    \begin{tabular}{c|c|c}
    \toprule
    \textbf{Index} & \textbf{Name} & \textbf{Description} \\
    \midrule
    1  & Pivot & It enables rotational movement around a single axis. \\

    2  & Ball-and-socket & It offers the widest range of motion, allowing for movement in all three planes, including rotation. \\

    3  & Hinge & It allows movement in one plane, similar to the way a door hinge works. \\

    4  & Condyloid & It allows movement in two planes: flexion and extension, as well as abduction and adduction. \\

    5  & Plane & It involves two flat surfaces that slide over each other, allowing for limited movement in multiple directions. \\

    6  & Saddle & It allows angular movements similar to Condyloid joints but with a greater range of motion. \\

    \bottomrule
  \end{tabular}
  }
\end{table*}

\begin{table}[tbh!]
  \centering
  \caption{\textbf{The metrics and the questions in the questionnaire}}
  ~\label{tab:questionnaire}
    \vspace{-0.3cm}
    \resizebox{\linewidth}{!}{
    \begin{tabular}{c|c|c}
    \toprule
    \textbf{Index} & \textbf{Metric} & \textbf{Question} \\
    \midrule
    
    1  & Easiness to use & I think this function is easy to use. \\

    2  & Learnability & I think I learned this function quickly. \\

    3  & Helpfulness & I think this function is helpful for realizing the IDI concept.\\

    4  & Expressiveness & I think this function is able to express my thinking for reconstruction. \\

    5  & Non-frustration & I think I can complete this function without frustration. \\
    \bottomrule
  \end{tabular}
  }
\end{table}

\end{document}